\documentclass[twocolumn,showpacs,preprintnumbers,amsmath,amssymb]{revtex4}

\def\pra{{Phys.~Rev.~A}}

\def\prl{{Phys.~Rev.~Lett.}}

\def\nat{{Nature}}

\def\jcp{{J.~Chem.~Phys.}}

\usepackage{isomath}

\usepackage{upgreek} 

\usepackage{color} 

\usepackage{graphicx}
\usepackage{mathtools}
\usepackage{dcolumn}
\usepackage{bm}
\usepackage{rotating}
\usepackage{amssymb}
\usepackage[latin1]{inputenc}
\begin{document}

\title{A Study of  molecular cooling via Sisyphus processes.}
\author{Daniel Comparat}
\affiliation{Laboratoire Aim\'{e} Cotton, CNRS, Universit\'{e} Paris-Sud, ENS Cachan, B\^{a}t. 505, 91405 Orsay, France}

\date{\today}

\begin{abstract}
We present a study of Sisyphus cooling of molecules: the scattering of a single-photon remove a substantial amount of the molecular kinetic energy and an optical pumping step allow to repeat the process.
 A review of the produced cold molecules so far indicates that the method can be implemented for most of them, making it a promising method able to produce a large sample of molecules at sub-mK temperature.
Considerations of the required experimental parameters, for instance the laser power and linewidth or the trap anisotropy and dimensionality, are given.
 Rate equations, as well as scattering and dipolar forces, are solved using Kinetic Monte Carlo methods for several lasers and several levels. For NH molecules, such detailed simulation predicts
a 1000-fold temperature reduction and an increase of the phase space density by a factor of 10$^7$ .
Even in the case of molecules with both low Franck-Condon coefficients and a non-closed pumping scheme, 60\% of trapped molecules can be cooled from 100 mK to sub-mK temperature in few seconds.
Additionally, these methods can be applied to continuously decelerate and cool a molecular beam \end{abstract}

\maketitle

\begin{figure}[ht]
\centering
\resizebox{0.4\textwidth}{!}{
		\rotatebox{-90}{
\includegraphics*[27mm,44mm][114mm,176mm]{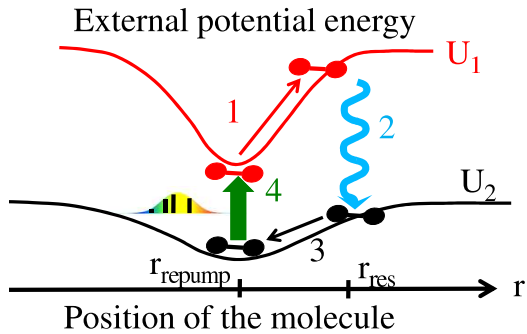}
		}
		}
		\caption{(Color online). Principle of Sisyphus cooling of molecules: 1) Removal of kinetic energy through motion in an external potential, 2) "Sisyphus transfer step": a dissipative process avoids the reverse motion. 3) This  ``one-way" (or ``single photon") process can be repeated by bringing the molecule back in position using the trapping potential and  (4) in its original internal state using light absorption in a "repumping step" (symbolized by the shaped spectrum for amplitude selection).
	 $r$ designs the radial coordinate. 	In  2D or 3D, due to the angular momentum the particles can miss the center, therefore $r_{\rm repump}$ could be non zero.
		}
\label{fig:opt_pump}
\end{figure}

In this article we study the Sisyphus method for cooling molecules. 
In such cooling,  sketched
 in Fig. \ref{fig:opt_pump},
external forces  remove kinetic energy by transferring it into potential energy.  A
 (absorption-)spontaneous emission step follows, creating non reversibility of the process which can be  repeated by optical pumping  the molecule back to its original state.
The method is first compared to other ones, then illustrated with a simple 1D or 3D model, and finally detailed on specific  cases, such as the NH molecule. We then conclude  that the method is very versatile and is able to produce
large samples of molecules at very low temperature.

\section{Introduction}

A first simple idea to 
 cool molecules is the use of evaporative cooling or  collisions with  dense and colder species, such as trapped laser cooled atoms or ions, in a so called sympathetic cooling scheme. Such a thermalisation technique has been demonstrated  with molecular ions  \cite{2000PhRvA..62a1401M}. 	For a long time, reactive or inelastic collisions have strongly limited the efficiency of the process for neutral species \cite{1981PhRvL..46..236D,2009FaDi..142..191S}. Only very recently OH radicals have been cooled using evaporative cooling \cite{2012Natur.492..396S}.

A second,  straightforward idea is the laser cooling technique  using
 transfer of  photon momentum.  While the first demonstration of light pressure on molecules occured in 1979 \cite{herrmann1979molecular}, it is
only very recently that laser cooling
 of molecules has been observed \cite{shuman2010laser}. To circumvent the general modification of the internal state occurring after the spontaneous emission step,
this has indeed required choosing a very well suited molecule (SrF), which  has a quasi-closed-level system with a very high Franck-Condon factor  (see also Table \ref{tab:mol}).

  A third route, suitable  for a larger number of molecular system, is to
modify the kinetic energy into potential energy using external forces \cite{1972JChPh..57.1487D}. 
A so-called one-way  \cite{2001PhRvA..64f3410B,2008PhRvL.100x0407T,2011PhRvL.106p3002F,2012PhRvA..85e3406S}, or irreversible, cooling 
 has to be realized in order to avoid the reverse process which otherwise would heat the sample.  
This  can be realized, for instance, by using 
an 
 (absorption-)spontaneous emission step \cite{2009NJPh...11e5046N,2010PhRvA..82b3419S}.
Finally, this can  be repeated, in a so-called Sisyphus cooling process  \cite{1985JOSAB...2.1707D} (see Fig. \ref{fig:opt_pump}).
Compared to standard laser cooling based on photon momentum transfer,  the reduction in temperature per spontaneous emission step is typically a few mK compared to  a few $\upmu$K. It can be a large fraction of the initial temperature. The process is therefore less sensitive to the modification of the internal state occurring after spontaneous emission.
The Sisyphus cooling process, first proposed by Pritchard \cite{1983PhRvL..51.1336P}, has represented a milestone in the
history of laser cooling by being responsible,  through polarization induced light shifts, for breaking the Doppler limit  in magneto-optical traps (MOT) 
  \cite{1985JOSAB...2.1707D}. The irreversible  Sisyphus  cooling cycle  has then been realized for atoms \cite{1986PhRvL..57.1688A} with the help of a gravity sag,  both in magnetic \cite{1995PhRvL..74.2196N} and
 in evanescent wave reflection fields \cite{1995OptCo.119..652S},  and with RF-induced transition both in an optical dipole trap \cite{2002PhRvA..66b3406M} and in a magnetic trap \cite{2005PhRvA..71a3422J}.
 A typical
experiment, as reported in Ref. \cite{2002PhRvA..66b3406M}, concluded that
``the final temperature achieved was (a somewhat disappointing) $17 \ \upmu$K". This explains why nowadays Sisyphus  cooling  is not heavily used by the cold atom community. But, such a low temperature would be a tremendous result for molecules. Indeed, a review, given by table \ref{lis_coldMolec}, of 
			the  produced  cold molecular species indicates that most of them are produced at temperature of hundreds of mK. 
The Sisyphus effect  has recently allowed the first efficient laser cooling of a polyatomic molecule
from 400mK down to 30 mK 
with the phase-space density increased
by a factor of 30  \cite{2012Natur.491..570Z}.
In such  molecular Sisyphus  cooling scheme  some specific  symmetric-top rotors, remain electrically trapped, and microwave transition between vibrational states provides the energy transfer
 \cite{2009PhRvA..80d1401Z}. Even if  limited to a specific type of molecule, this extremely promising experiment already indicates the possibility to reduce the temperature  below the mK range in tens or hundreds of seconds.
One of the difficulties is to find molecules with 
 short enough spontaneous emission times. A similar problem is found in
Ref. \cite{2009JPhB...42s5301R}, where it is suggested to put an optical cavity around a magnetically trapped OH sample in order to accelerate the spontaneous decay.

In this article,
we suggest to use optical pumping and faster electronic transitions in order to generalize the method. We first discuss the required ingredients needed for efficient cooling such as the choice of the molecule, the trap and laser parameters. We then perform a detailed simulation of the whole process in realistic conditions in the case of NH molecule. 

\section{COOLING STRATEGY}

\subsection{Choice of the molecular states}

	Each molecule has is own characteristics and the cooling strategy should be adapted to each of them
	by making choices for the trap, the  electronic states and the lasers.

			In Table \ref{tab:mol}, we have listed some diatomic molecules \footnote{Parameters have been taken from the webbook.nist.gov website and the Frank-Condon between the lowest vibrational state has been estimated from a Morse potential calculation from http://www.yale.edu/demillegroup/sharedfiles/Franck Condon Symplectic Integrator.nb.
			A simpler harmonic approximation for the potential curves using the reduced mass $\mu$ and the given $\omega_e'', \omega_e'$ $r_e'$ and $r_e''$, leads for the Franck-Condon  (in SI units) $2\frac{\sqrt{\omega_e'' \omega_e'}}{\omega_e' + \omega_e''} e^{ - \mu (r_e' - r_e'')^2   \omega_e'' \omega_e' / \hbar (\omega_e' + \omega_e'')}$.} to illustrate their properties and explain why we think that 
			the method can be implemented for most of the up to now    produced  cold molecular species.
			 For simplicity, in this article, we do not consider any hyperfine structure, the  splitting of which being typically in the MHz range and if needed can be resolved and manipulated by  laser sidebands \cite{shuman2010laser}. We simply note that the role of the hyperfine structure can be very important and, contrary to common thought, it can make the cooling simpler; for instance by  creating the two trapping potentials $U_1, U_2$ when without the hyperfine structure only a single one would exist.

			For the choice of the lasers or trap, a key parameter is the AC or DC electric, magnetic or electromagnetic
 trapping capability of the molecule.

Electrostatic or magnetic trap can be used if the molecule present a strong enough energy shift $\hbar \Delta $ due to Stark or Zeeman effect.
   In an external electric  $\bm E$ or  magnetic $\bm B$ field, the energy shift  of the $  {}^{2S+1} |\Uplambda|_{\Upomega}$ state depends on the  rotational   level $J$.  For Hund's case a electronic state, it is
  respectively $ E d \Omega  M /J(J+1)$ and
 $B \mu_{\rm B}    (  \Lambda - 2 \Omega) \Omega M /J(J+1)$ 
 where $\mu_B$ is the Bohr magneton, $d$ is the permanent electric dipole moment and
 $ M$  the projection of $\bm J/\hbar$ along the  quantization axis  given by the local  field  at the particle position. Most of the molecules listed in Table \ref{tab:mol} can thus be trapped in electric or magnetic trap. 
However some of them (with  $^1\Upsigma$ ground state for instance) 
 cannot be electrically or magnetically trapped.
In such case, one solution is to transfer the molecule to a state with sufficiently long lifetime where the method can then be applied  \cite{1999PhRvL..83.1558B,2011PhRvL.106u3001W}.
Another solution, is to use 
 a  laser dipole  trap \cite{1986PhRvL..57.1688A,2011PhRvA..84f3417I}. But,
because the initial temperature of the molecule is usually high (cf table \ref{lis_coldMolec}) a quite deep trap is required and so a 
close to resonance laser has to be used.
 For instance a 100 W laser focused on 1 mm diameter, detuned a fraction of a nm from a resonance leads to a typical trap depth of more than 10 mK  and an oscillation frequency of $\omega \approx 2 \uppi (500 $ Hz$)$.  Such laser can be realized through (fiber) laser amplification or using a Quasi-Continuous Wave laser \cite{2009ApPhB..94...71C}.
Interestingly enough, the fact that the laser is close to resonance can be used to take benefit of the high diffusion rate ($10^{5}$ photons/s in the previous example) to perform the Sisyphus transfer between the two trapped state \cite{1986PhRvL..57.1688A}. 			
			
			\begin{table}[htbp]
					\centering
	\begin{tabular}{|l|l|l|l|l|l|}
	\hline\noalign{\smallskip}
Molecule &	Low state  & High state  & $\lambda$ (nm) & FC & lifetime \\
	\hline\noalign{\smallskip}
		\hline\noalign{\smallskip}
		NH & X $^3 \Upsigma^-$  &  X $^3 \Upsigma^-$ & 3050 & 1 & 37 ms \\
		 & X $^3 \Upsigma^-$  &  A $^3 \Uppi$ & 336 & 1 & 400 ns \\
					&  X $^3 \Upsigma^-$ &  b $^1 \Upsigma^+$ & 471 & 1 & 18 ms \\
				 &  X $^3 \Upsigma^-$ &  a $^1 \Updelta$ & 794 &  1 &  12 s \\
					 {\it BH} & X $^1 \Upsigma^+$  &  A $^1 \Uppi$ & 433 &  1 &  160 ns \\
					 YO & X $^2 \Upsigma^+$  &  A $^2 \Uppi$ & 614 &  0.99 &   \\
				{\it CH} & X$^2 \Uppi$  & A$^2 \Updelta$ & 431 & 0.99 & 530 ns \\
		SrF & X $^2 \Upsigma^+$  &  A $^2 \Uppi$ & 651 & 0.98  & 23 ns \\
			CaH 	& X $^2 \Upsigma^+$  &  A $^2 \Uppi$ & 693 &  0.98 &  \\
	CaF 	& X $^2 \Upsigma^+$  &  A $^2 \Uppi$ & 606 & 0.98 & 20 ns \\
			YbF & X $^2 \Upsigma^+$  &  A $^2 \Uppi$ & 552 &  0.95 &  \\
				BaF 	& X $^2 \Upsigma^+$  &  A $^2 \Uppi$ & 860 &  0.95 & 50 ns  \\
				{\it MgH} & X $^2 \Upsigma$  &  A $^2 \Uppi$ & 519 &   0.94 &  \\
				YbF 	& X $^2 \Upsigma^+$  &  A $^2 \Uppi$ & 552 &  0.93 & \\
			OH & X $^2 \Uppi$  &  A $^2 \Upsigma^+$ & 308  & 0.91 & 700 ns \\
					ThO & X $^1 \Upsigma^+$  &  A $^1 \Upsigma^+$ & 943 & 0.86 & \\ 
			VO & X $^4 \Upsigma^-$  &  B $^4 \Uppi$ & 787 &  0.80 & \\
		MnH & X $^7 \Upsigma$  &  A $^7 \Uppi$ & 556 &  0.78 & \\
			{\it CS}  & X $^1 \Upsigma^+$ & A $^1 \Uppi$ & 258 & 0.78 & 200 ns\\
	{\it TiO}		& X $^3 \Updelta$  &  A $^3 \Upphi$ & 709 &  0.72 & \\
		CrH & X $^6 \Upsigma^+$  &  A $^6 \Upsigma^+$ & 866 & 0.68 & \\	
			{\it CaS}  & X $^1 \Upsigma^+$  &  A $^1 \Upsigma^+$ & 658 & 0.59 &  \\
		{\it CN}  & X $^2 \Upsigma^+$  &  A $^2 \Uppi$ & 1097 & 0.50 & 10 $\upmu$s \\
		{\it CaO}  & X $^1 \Upsigma^+$  &  A $^1 \Upsigma^+$ & 866 & 0.43 &  \\
		 & X $^1 \Upsigma^+$  &  A' $^1 \Uppi$ & 1199 & 0 &  \\
		CO  & X $^1 \Upsigma^+$  &  a $^3 \Uppi$ & 206 & 0.31 & 10 ms \\
			     & X $^1 \Upsigma^+$  &  A $^1 \Uppi$ & 154 & 0.12 & 10 ns \\
					& a $^3 \Uppi$   &  a' $^3 \Upsigma^+$  & 1453  & 0.04 & 3 $\upmu$s\\
						& a $^3 \Uppi$   &  d $^3 \Updelta$  & 823  & 0.02 & \\
	SrO 	& X $^1 \Upsigma^+$  &  A $^1 \Uppi$ & 920 &  0.29 & \\
			 & X $^1 \Upsigma^+$  &  A' $^1 \Uppi$ & 1074 & 0 &  \\
			{\it SO } & X $^3 \Upsigma^-$  &  A $^3 \Uppi$ & 262 & 0.22 & 12 ns \\
		NO & X $^2 \Uppi$  &  A $^2 \Upsigma^+$ & 227 & 0.17 & 200 ns\\
				& X $^2 \Uppi$  &  a $^4 \Uppi$ & 263 & ? & 100 ms\\
				{\it SiO } & X $^1 \Upsigma^+$  &  A $^1 \Uppi$ & 235 & 0.15 & 9 ns \\ 
		{\it N$_2$} & X $^1 \Upsigma_{\rm g}^+$  &  a $^1 \Uppi$ &  &  0.04 &   \\
					 & A $^3 \Upsigma_{\rm g}^+$  &  B $^3 \Uppi_{\rm g}$ &  &   & 8 $\upmu$s \\
					PbO & X $^1 \Upsigma^+$  &  a $^3 \Upsigma^+$ & 629 &  0.02 & \\
		H$_2$ & X $^1 \Upsigma_{\rm g}$  &   B $^1 \Upsigma_{\rm g}^+$ & 111 & 0.01  &  1 ns \\
				 & c $^3 \Uppi_{\rm g}$  &   j $^3 \Updelta_{\rm g}$ & 567 & 0.99  &   \\
				LiH & X $^1 \Upsigma^+$  &  A $^1 \Upsigma^+$ & 385 &  0 & \\
     & X $^1 \Upsigma^+$  &  B $^1 \Uppi$ & 291 &  0.08 & \\
				 O$_2$ & X $^3 \Upsigma_{\rm g}^-$  &   B $^3 \Updelta_{\rm g}^-$ & 286 &  0 & predissociation \\
				\hline\noalign{\smallskip}
\end{tabular}
				\caption{Some properties of some diatomic molecules. In italic are the molecules that have not been cooled up to now (cf. table \ref{lis_coldMolec}).
				The molecules are ordered by the rounded value of the  $v'=0,v''=0$ Franck-Condon factor.	The main parameters needed to realize the Sisyphus cooling is a trapping capability (depending on the electronic state), a transition wavelength in the region cover by laser (typically near the visible region) and a good Franck-Condon ratio in order to simplify the repumping step. The spontaneous emission (lifetime) is mainly here to ensure the feasibility of the process.}
				\label{tab:mol}
			\end{table}

\subsection{Removal elementary step strategy}

Once the trap chosen, the cooling strategy is strongly linked to the energy removal elementary steps: step 1 and 2 in Figure \ref{fig:opt_pump}.

	In order to study the dynamics, we shall assume
 $N$
particles initially at temperature $T$ in potential  
 $U_1(r)$. They are then transferred to the potential  $U_2(r)$ and then, when reaching $r_{\rm repump} =0$ (in the 1D case), are pumped back to $U_1$. 
For simplicity, we shall illustrate first the strategies in the simple 1D picture and using two harmonic potentials $U_1(r) =m \omega_1^2 r^2 /2$ and $U_2(r) =m \omega_2^2 r^2 /2$ (where gravity is neglected). 
	In such process there are no collisions to thermalize the sample, but for simplicity we shall use an effective temperature $T$ (given by the average kinetic energy). 

\begin{figure}[ht]
\centering
\resizebox{0.5\textwidth}{!}{
		\rotatebox{-90}{
\includegraphics*[28mm,52mm][170mm,250mm]{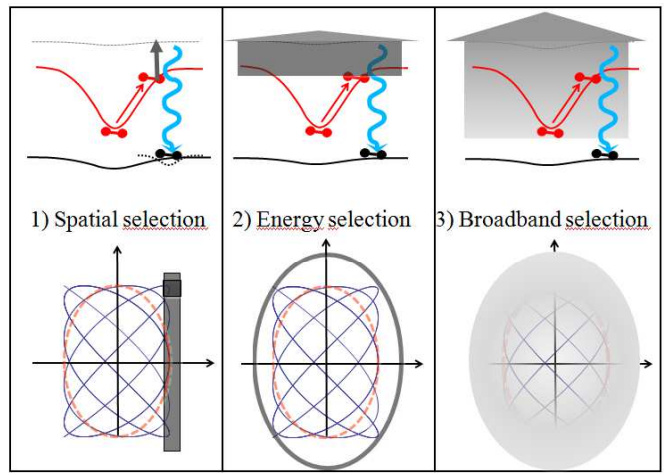}
		}
		}
		\caption{(Color online). Illustration of possible strategies to remove the kinetic energy from the sample. The upper part illustrate the process in a 1D picture whereas the lower part indicates it in a 2D picture. The transfer position is indicated by an hatched area. Two kinds of trajectories, but with the same kinetic energy, are shown: the dashed red line correspond to an isotropic trap case $\omega_x = \omega_y$ and the solid line with $\omega_x = 0.8 \omega_y$. 1) Left panel: the selection is done spatially (in the hatched region shown by the line or box) to catch the molecule where it has almost zero kinetic energy. 2) Central panel: the selection is done spectrally depending on the potential energy of the particle  when it has almost zero kinetic energy. 3) right panel: the selection is done depending on the velocity through the lifetime of the (absorption or spontaneous emission) transition.}
\label{fig_remov_energy}
\end{figure}

The Sisyphus transfer between state $1$ and $2$ can be obtained is several ways and using
different strategies illustrated by \ref{fig_remov_energy}  such as by Spontaneous or induced transfer that we want to discuss briefly:

\begin{itemize}

\item Spontaneous emission (figure	 \ref{fig:opt_pump})

As shown by the figure	 \ref{fig:opt_pump}, the simplest solution is simply to wait for spontaneous emission.
	This can be used when the first state ($U_1$) has a long enough lifetime so that the particle  moves in the trap before its spontaneous decay
 \cite{1994OptCo.106..202H,2006PhRvL..97u3001P,2006JPhB...39.4945T}.

Unfortunately, in many cases $U_1$ and $U_2$ will simply be two rotational or vibrational states of the electronical ground state having a very long spontaneous decay time. For such cases, another way is to control  the Sisyphus step by applying an excitation laser,  or any other light source (RF-microwave, ...), to the molecule initially in the potential $U_1$ to excited it to an excited state that will then decay towards $U_2$.

\item Continuous transfer (right and upper part of Figure \ref{fig_remov_energy}):

A simple way to mimic this case of a spontaneous  emission  with rate $\gamma$ is to
realize the Sisyphus transfer step by using a large and broadband laser covering the  whole sample.
 This creates  a quite uniform excitation rate $\gamma$ toward an excited state that (quickly) decay toward $U_2$. Usually $\gamma$ is  chosen to be smaller or on the same order of magnitude  than the trap oscillation frequencies in order to avoid too fast transfer that would transfer the molecule before it has lost enough energy.

In such process (spontaneous emission or uniform excitation), in average, the particle  loose kinetic energy simply by the fact that 
a slowing down molecule spends more time at the spacial points where the velocity is reduced. We can draw here some similarity with the optical pumping scheme in  a blue optical molasses where atoms scatter less photons when they have less kinetic energy because there is less light present
\cite{1989JOSAB...6.2023D,ISI:000085557900004}. 
Hence, under the effect of its motion the molecule will be transferred toward $U_2$ when it has the smallest kinetic energy. 
More precisely, neglecting the photon recoil energies, and using survival probability described in the appendix \ref{KMC_model},  we see that the average energy reduction is given by $\int_0^{+\infty} (U_1(r(t)) - U_2(r(t)))\gamma e^{-\gamma t} {\rm d} t$.
	Thus, in our simple harmonic trap example, using the typical trajectory $r(t)$ for the average kinetic energy $k_{\rm B} T$, 
	this energy reduction is $k_{\rm B} \Delta T = 0.4 k_{\rm B} T (1-\omega_2^2/\omega_1^2)$ for $\gamma = \omega_2$  \cite{2011PhRvA..84f3417I}.
 This value can be easily improved by using a non uniform excitation rate $\gamma(r)$ as graphically illustrated by the case 3 in Fig. \ref{fig_remov_energy}.
Moreover, we can always
combine these ideas and, depending on the oscillation frequencies, laser power and  cooling time available, find the  optimum spatial distribution of the  light power which optimizes the one way cooling.

\item Induced spatial transfer (left and upper part of Figure \ref{fig_remov_energy}):

When using a laser induced transfer,
one  idea is not to cover the whole sample by laser, but to catch the  particle at the "top" of its trajectory that is when it has almost zero kinetic energy. 
 This selection can be realized spatially by applying the laser only at this position. This simple realization is illustrated by the localized arrow in the left and upper part of Figure \ref{fig_remov_energy}.

\item Induced energy transfer  (cf. middle and upper part of Figure \ref{fig_remov_energy}):

However, it can be sometimes simpler to apply the laser on the overall sample and to realize this selection spectroscopically. We can use the fact that the resonance condition is fulfilled only locally due to different potentials seen by the particles during the excitation (case 2 of Figure \ref{fig_remov_energy}).

For this last two cases, we can assume that
  the laser transition occurs only at location $r= r_{\rm res}$.
Due to Boltzmann statistic the number of molecules  reaching the transition point  are $N {\rm e}^{-  \eta}$, where
  $\eta = U_1(r_{\rm res})/k_{\rm B}T$. 
 Assuming that the particles hardly move before being optically (re-)pumped, i.e. that 
all (re)pumping rate are faster than the trap frequencies, 
	 the total  energy change 
for the sample is then $ \Delta	 E = N k \Delta T = N {\rm e}^{-  \eta} (U_2(r_{\rm res}) - U_1(r_{\rm res})) <0 $. 
The molecule has spent at most a quarter of the oscillation time in each potential before being transfer to the other one. Thus,	
	a maximum cycling time is $\Delta t = \pi/2 \omega_1 + \pi/2 \omega_2$.
This lead to the equation
\begin{equation}
\frac{\Delta T }{\Delta t}  = - \frac{2 \omega_1  \eta {\rm e}^{-\eta} }{\pi } \frac{1-\omega_2^2/\omega_1^2}{1+\omega_1/\omega_2} T  
\label{eq_T_1D}
\end{equation}
 As for evaporative cooling processes \cite{Ketterle1996b,2006PhRvA..73d3410C},   
we see that
it is probably efficient to adjust the laser detuning (or the external trapping potential) as the particles  are losing energy. This can be done, for instance, by keeping the ratio $\eta $  of the transition energy to the thermal energy constant, with $\eta = 1 $ being the best choice.
 The speed can  be optimized by choosing $\omega_2 = \omega_1/2$  with a temperature variation per cycle $\Delta T \approx 0.28 T$ and a (1/e) decay time of $ 17 /\omega_1 $.
\end{itemize}

Independently of the chosen strategy we can drive the conclusion that a 1D Sisyphus cooling will provide
a temperature variation per cycle on the order of $\Delta T \approx 0.3 T$. Then
the temperature should decays exponentially with a (1/e) decay time on the order of $ 20 /\omega_1 $.
Thus, cooling by a factor 10000, say from 100 mK to 10 $\upmu$K, takes only 30 cycles. 
 This has to be compared to the few thousands of cycles needed for standard laser cooling \cite{shuman2010laser}. 
This simple study clearly highlights the potential of the Sisyphus method.

\subsection{1D, 2D or 3D cooling?}

In real the particles moved in three dimension (3D)
 for instance being trapped in 
an harmonic trap with
$U({\bm r}) = \frac{1}{2} m \omega_x^2 x^2 + \frac{1}{2} m \omega_y^2 y^2 +\frac{1}{2} m \omega_z^2 z^2$. 
The problem is that, even in such simple case,
 where the motion in one of the three axis x,y,z is decoupled from the others, 
the previous discussion, concerning 1D motion, cannot be generalized to the (2D or) 3D case. 
The reason is that, if Fig. \ref{fig:opt_pump} is still valid, 	where $r$ designing the radial coordinate, now
 in 2D or 3D motion, due to the angular momentum, the particles can miss the center  and $r_{\rm repump}$ could be non zero.
This is clearly indicated by the trajectories, calculated in the case of an harmonic  2D potential and shown in the
lower 
  part of Fig. \ref{fig_remov_energy}. The dashed red line trajectory, corresponding to an isotropic trap case $\omega_x = \omega_y$, performs at constant $r$. Thus the process given in Fig.  \ref{fig:opt_pump} does not work at all because $r$ does not change, so the trajectory never crosses the desired transfer point (box in case 1 and solid circle in case 2 of Fig. \ref{fig_remov_energy}) where the kinetic energy should be zero, simply because the kinetic energy is constant and never zero. 
	
	There is at least two ways to go around this important difficulty:

	\begin{itemize}
		\item Create asymmetries in the potential
		
In order to avoid particles that orbit around the trap center with constant $r$ for instance,
it is possible to
 create asymmetries in the potential \cite{1994OptCo.106..202H,1995PhRvL..74.2196N,2009JPhB...42s5301R,2009NJPh...11f3044T}.
This is illustrated by the second trajectory (solid line in the lower 
  part of Fig. \ref{fig_remov_energy}), corresponding to an anisotropic trap case $\omega_x = 0.8 \omega_y$. The trajectory is more complex now and present some points where the kinetic energy is almost zero and thus the trajectory crosses the line or box where the transfer should occur and the cooling will be efficient.

An  important message is that a small modification of the trap can totally modify the efficiency of the cooling (see also Fig. \ref{fig:aniso_trap}).

\item $3\times1$D cooling

If we only look on one axis (let say the $x$ axis), the 1D Sisyphus cooling picture given previously is perfectly valid and efficient with $r=x$. Indeed, the motion along x crosses the zero center and, at the edge of the trajectory, the velocity along x is zero as desired for an efficient Sisyphus transfer.
The usual 1D process can thus be used to reduce first the velocity, or the temperature $T_x$, along this axis. We then just have to repeat the process for the other axis y and z. Experimentally, the cooling along x can be done by using a spatial selection that affects only the $x$ coordinate but not the other ones. For this we can choose the induced spatial transfer (case 1) strategy and  send the laser beam only along a line (or a plane in a 3D vision) as shown on the left lower part of
 Fig. \ref{fig_remov_energy} by the hatched area.
	\end{itemize}

		\subsection{Phase space density and optimized strategy}

The temperature is not the only relevant parameter. Another important parameter, which is to be maximized, is the phase space density $D = n_0 \Lambda^3 $, where $n_0$ is the peak density
and $\Lambda = \sqrt{\frac{2  \pi \hbar^2 }{m k_B T}}$ is the thermal de Broglie wavelength  \cite{2006PhRvA..73d3410C}. 
In the 3D isotropic harmonic case, for $U(r) =\frac{1}{2} m \omega^2 r^2  $, we have $D= N \left( \frac{\hbar \omega}{k_B T} \right)^3$ 
and $n_0 = \frac{N}{(2 \pi \sigma_r^2)^{3/2}} $  where $U(\sigma_r)  = \frac{1}{2} k_B T $.

A proper optimization of the cooling should be done by clearly defining the
objective function to be optimized. 
Depending of the goal, we might thus  prefer to optimize the time, the number of particles, the temperature, the phase space density or, as done experimentally for evaporative cooling strategies, the parameter
 $\frac{\dot D/D}{\dot N/N} $ \cite{Ketterle1996,1997PhRvA..55.3797S,2006PhRvA..73d3410C}. The strategy has also to be chosen depending of constrains such as: vacuum limitation,  available lasers, trapping possibilities, ...

We have seen that 
the efficiency of the process strongly depends on its dimensionality, on the trap anisotropy and on the transfer strategy chosen. Comparison of all possible processes and dependance with trap or laser parameters is beyond the scope of this article. However, we have tried to initiate such discussion in the Appendix \ref{cool:strategy} and the results are summarized by the Fig. \ref{tab:scaling} which gives the
time evolution of number of molecules $N$, Temperature $T$ and phase space density $D$ under different cooling strategies.
These results should be taken only as rule of thumbs and not as accurate predictions. In fact, one  important consideration is  that the interaction  time, the power broadening of the transition or the velocity dependance of the transition rate can lead to a much
 lower energy or spatial resolution than naively expected. We thus need to confirm these results by simulations. But, before doing so 
we first extract, from the appendix, few considerations for the efficiency for three different cooling strategies:

\begin{itemize}

\item one-way or single photon cooling.

In the first step (1 and 2) of Fig. \ref{fig:opt_pump},  a single photon cooling  decreases the temperature but the density stay the same because we simply remove the kinetic energy. However,
it is possible to increase furthermore the phase space density by transferring the particles in a tighter trap without heating.
 This can be done by
  catching them in a tight trap as indicated by  the dashed trap shown in the upper panel of case 1 of Fig. \ref{fig_remov_energy}).
In order to catch all particles  
it is even possible to have this  small, tight  trap $U_2$ (the black box in the lower panel of the figure) that  dynamically follows the location of the transfer.
In principle such single photon cooling processes looks extremely attractive \cite{2011arXiv1112.0916S}: starting with the more energetic
molecules, and slowly sweeping the laser position (for case 1) or the transition frequency (for case 2) down,  we can always transfer the molecules when they have almost zero kinetic energy. A simple optimization of the time evolution (trajectory) of this box is proposed in Ref. \cite{2012arXiv1204.0352S}. 
If the lower potential $U_2$ is almost flat we can also in principle remove all the energy with a single photon emitted  per molecule.
Several realizations  of this one-way or single photon cooling already exist, but only for atoms \cite{2008PhRvL.100x0407T,2009NJPh...11e5046N,2009NJPh...11f3044T,2010PhRvA..82b3419S,2011PhRvL.106p3002F,2011EPJD...65..161R}.

	These one-way or single photon cooling schemes has to be used if the repumping step 4) of Fig. \ref{fig:opt_pump} is difficult to realize. An obvious example is when the decay after step 2) populate a lot of levels, for instance due to bad Frank-Condon factors, because it will be quite difficult to optically pump all of them.

\item $3\times1$D cooling

The $3\times1$D cooling is very efficient because it is simply three times the 1D cooling one. For instance the time needed to reach the same final temperature  has to be multiplied by a factor $3$ compared to the 1D case. It is thus a very fast cooling and it works with any kind of trap geometry.
Another advantage of such 1D cooling, compared to the 3D one, is that the quantification axis is always the same (there is only one possible axis!) and transitions can thus be controlled using light polarization, which can be extremely useful to excite only the states which are trapped (for instance by controlling the Zeeman sub-level in a magnetic trap). A last advantage is that the particles always cross the 0 coordinate along the considered axis and so the repumping light can then be focused only at the center of the trap in order to realize the ideal step 4 of Fig. \ref{fig:opt_pump}.
Finally we would like to mention the very useful rotation of phase space method used in Ref. \cite{1995PhRvL..74.2196N} in order to cool always along the same direction. Quoting the authors:
``First the vertical distribution is cooled. Then, by adiabatically changing the trap parameters, we produce a degeneracy $\omega_x \sim \omega_z$. Anharmonicities cause the atom cloud to rotate in the x-z plane, exchanging the x and z distribution. After we adiabatically restore the trap to the initial field conditions. We then cool the vertical distribution again."
Therefore only one laser direction is enough to cool the 3D sample!

The conclusion is that this three times 1D cooling may be simpler and more efficient that direct 3D cooling \cite{1995PhRvL..74.2196N,2009PhRvA..80d1401Z} but its practical implementation has to be done with care since the other axis should be spectator, not only in the cooling step but also in the repumping step.

	\item Standard 3D Sisyphus cooling scheme.
	
	In the following we have chosen  to study in more detail the Sisyphus cooling scheme because it keeps all molecules and does not require to move lasers. But , even without a detail simulation, we can directly see from Fig. \ref{fig:opt_pump} that, as the particle returned to the original potential   (step 3), the reduction of temperature  also induces an increase of the density and of the phase space density.

\end{itemize}

\section{Sisyphus cooling on NH: Case with high Franck Condon value $FC \sim$ 1.}

To catch more insight of the Sisyphus cooling process, we shall first study a case where the repumping step is simple. 
In order to have a fast cooling  process we want to use an electronic transition.
The simplest choice it to choose molecule with
a Franck Condon factor close to unity creating a closed vibrational pumping scheme.
With such close cycling scheme the Sisyphus cooling process is clearly advantageous compare to the one-way single-photon one because we can repeat the single photon process without losses.

As an example we shall investigate the
	NH (Imidogen) molecule. The A-X
transition in NH is indeed almost perfectly diagonal and the
v' = 0 - v'' = 0 band has a Franck-Condon factor of better
than 0.999 \cite{2011EPJD...65..161R}. NH is a very well known molecule which
has already been decelerated and cooled  (see table \ref{lis_coldMolec}). It has been trapped both in electric \cite{2007PhRvA..76f3408H} and magnetic traps, for more than 20 s \cite{2010NJPh...12f5028T}, in the a$^1 \Updelta$ or its X$^3 \Upsigma^-$ state respectively.
The accumulation of Stark-decelerated NH molecules in a magnetic trap has been realized \cite{2011EPJD...65..161R}.

 NH is probably one of the simplest systems to study cooling of molecule.
A 	single photon cooling scheme has been envisioned  \cite{2009NJPh...11e5046N}.
For simplicity we shall concentrate here on NH in its X state trapped in magnetic field. But the case of the a$^1 \Updelta$ state and an electrostatic trap is also possible with
the 325 nm
 c$^1 \Uppi \leftarrow $a$^1 \Updelta$ Sisyphus transition and  repumping on the 452 nm c$^1 \Uppi \leftarrow $b$^1 \Upsigma^+ $ transistion.

A 471 nm  
b$^1 \Upsigma^+ \leftarrow $X$^3 \Upsigma^-  $ transition is possible but a simple and efficient transition is the 
 A$^3 \Uppi; v' = 0 \leftarrow $X$^3 \Upsigma^- ; v'' = 0$ transition.

In order to perform the Sisyphus cooling we  need to find two states with different magnetic moment and then couple them using  laser light through an intermediate level.
 We choose not to use the absolute rotational ground state but the X$^3 \Upsigma^- ; N'' = 1, J''=1 +$. This allows to
  perform the Sisyphus cooling on the  A$^3 \Uppi; N' = 1, J'=0, - \leftarrow $X$^3 \Upsigma^- ; N'' = 1, J''=1 +$ transition that is,  due to parity reasons, the only closed $ N'  \leftarrow N'' $ rotational transition.
A sketch of the relevant levels and transition strength, is shown in Figure \ref{fig:NH_X} as calculated by a Program for Simulating Rotational Structure (PGOPHER) \footnote{A Program for Simulating Rotational Structure, C. M. Western, University of Bristol, http://pgopher.chm.bris.ac.uk.} with the 	 molecular constants taken from Ref.
	\cite{2010JMoSp.260..115R}. 
	From  Fig. \ref{fig:NH_X}, we can find two states with different magnetic moments to perform the Sisyphus cooling: $U_2$ being the  potential of the $M''=1$ level (with a linear Zeeman shift) and $U_1$ being the $M''=0$ one (with a quadratic Zeeman shift). One drawback is the possible decay towards the untrapped $M''=-1$ level, but, because of the fast electronic transition, a fast repumping from this state seems nevertheless possible.

		Even if it is not the purpose of this article, we mention that using this transition standard laser cooling of NH should be quite easy.
			The fact that the transition is a $J'' \rightarrow J'= J''-1$ scheme and not the standard  $J \rightarrow J + 1$ one for magneto-optical trap (MOT), may lead to some difficulties because in 1D this produces a dark state that prevents the cooling \cite{2008PhRvL.101x3002S,shuman2010laser}. However the dark state is not dark for the other lasers present in a 3D MOT.
		We have simulated such 3D cooling and found results comparable to the experimental demonstration  
of the so called 3D
	type-II magneto-optical trap  on sodium atom
	 \cite{1997PhRvA..55.4621O}. 
	However, as stressed in the Appendix, rate equations are not able to deal correctly with dark states so the simulation of a $J'' \rightarrow J'= J''-1$ transition using rate equation may be questionable.

\subsection{Cooling in quadrupolar trap with energy degeneracy}
	
	For the Sisyphus cooling, we first study the simple case of a quadrupolar trap. We explain in detail our choice for the parameters in a way that can be generalized for other situations and then present the results.
	
\begin{figure}[ht]
\centering
\resizebox{0.5\textwidth}{!}{
		\rotatebox{-90}{
\includegraphics*[29mm,40mm][200mm,270mm]{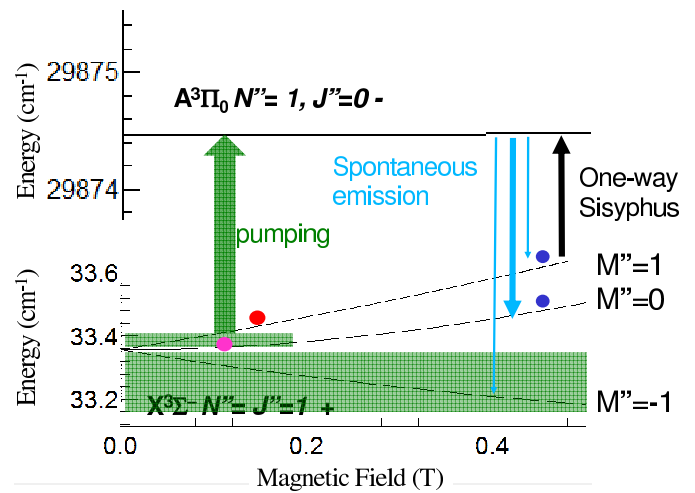}
		}
		}
		\caption{(Color online). Principle of NH Sisyphus cooling of molecules. The width of the arrows  are proportional to the branching ratios. }
\label{fig:NH_X}
\end{figure}

The magnetic field is thus
$\bm B (\bm R) = B'_0(1+\alpha)  X \bm e_{\rm X} + B'_0(1-\alpha)  Y \bm e_{\rm Y}) - 2 B'_0 Z \bm e_{\rm Z} $ 
with  $B'_0=100 $ T/m and $\alpha $ is a parameter that controls the anisotropicity of the trap.  We choose the anisotropy parameter $\alpha=0.17$ (see figure \ref{fig:aniso_trap}) which optimize, after 20 ms, both the energy conversion for the pumping and repumping steps.
	 This situation can  easily be experimentally realized, for example using 2 orthogonal pairs of coils  in  anti-Helmholtz configuration. 
In a gradient of $\sim 100 $ T/m 
	a typical molecule with potential energy of $\sim 150$ mK (corresponds to 0.1 cm$^{-1}$), sees a magnetic field slightly less than 0.2 T (see Fig. \ref{fig:NH_X}). This leads to a  typical radius for the sample of $\sigma_r \approx 2 $ mm. 	
	In order to affect most (say 90\%) of the molecules, we choose an initial laser waist size of 5 mm and	 we start
	the Sisyphus process
	with a laser detuning corresponding to a transition with a potential energy of four times the energy: 500 mK (in temperature units).

The time evolution of the cooling is linked to two parameters: the typical "oscillation" time $T_1$ (2 ms in our case) and  the number  of "oscillations" required before the molecule is transferred. As indicated in the appendix, this number  $N_0^{d-1}$ (1 for a 1D,  $N_0$ for a 2D, and $N_0^2$ for a 3D cooling) is linked to the selectivity $\Delta E$ in energy  (or in spatial position) of the transition by  $\Delta E  \approx E/N_0 $ in the 3D case.
 In our case,  
the study of the trajectories in the trap (see energies in figure \ref{fig:aniso_trap}) indicates that an accuracy on the order of $\Delta E  = 20-70$ mK is probably a good choice for a transfer after 20 ms. This suggests to take a  value 
$N_0 < 5$ for our case of an initial average potential energy of $\sim 150$ mK (temperature of 100 mK). Thus, we  expect, a  typical decay time for the temperature on the order of  $N_0^2 T_1$, i.e. below 50 ms.  
 We thus choose in our simulation to reduce all lasers powers, waists and detunings (i.e. transition energy) exponentially with a time constant of $\tau = 50$ ms.

The choice of the laser power is discussed in the Appendix \ref{param_need} and the results is that the intensity  $I = \frac{2 P}{\uppi w^2}$ should be on the order of $ \frac{k{\rm _B} T}{ \hbar \Gamma} \frac{v}{\sigma_r} \frac{\Gamma_{\rm L}}{\Gamma} \frac{(500 {\rm nm})^3}{\lambda^3}
  \frac{1}{300 \, 000 \, \rm FC.AF }    $.
For transition at $\lambda = 336$ nm with a linewidth $\Gamma = 1/(400 {\rm ns})$ (see table \ref{tab:mol}), a Franck Condon factor FC=1, an angular factor $AF= 1/4$ (ratio of the worst transition to the total one in Fig. \ref{fig:NH_X}), and a velocity $v$ at the transition taken to be typicality (for reasonable $N_0$ value) tens of the thermal velocity; a standard $\Gamma_{\rm L} = 2 \uppi (10 MHz)$ linewidth laser requires a power of nearly 100 mW.  We choose such values for the simulation.

For this first study, but this will be refined later, we do not put any realistic repumping laser but we choose a constant and uniform repumping rate of $10^6$ s$^{-1}$ for all molecules having a potential energy bellow 15 mK as indicated by the green box in Fig. \ref{fig:NH_X}.

For the simulation, the  absorption-emission processes are  accurately calculated using rate equations  
   taking into account the saturation, the laser detuning its linewidth and the Doppler effects \cite{Woh1}.  
The simulation used
a Kinetic Monte Carlo algorithm, to solve exactly the
 rate equations,  and  a Verlet algorithm to drive the particles motion under the  effects of gravity,  magnetic field 
 and recoil photons \cite{2008NJPh...10d5031C}. The molecular dipole moment is assumed to adiabatically follow the local field and the laser polarization is thus calculated along this local axis. Details are given in the Appendix, mainly by equation (\ref{eq_rate}). 

\begin{figure}[ht]
\centering
\resizebox{0.4\textwidth}{!}{
		\rotatebox{-90}{
\includegraphics*[13mm,28mm][191mm,269mm]{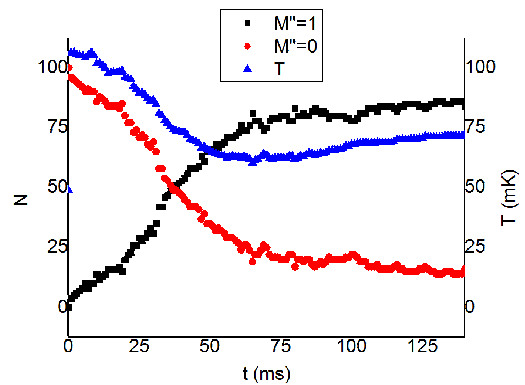}
		}
		}
		\caption{(Color online). Results of NH Sisyphus cooling of molecules in a linear (quadrupole) trap with energy degeneracy at center. Molecules number $N$ (left axis) in the $M''=1$ (square) and $M''=0$ (circle) states as well as the temperature $T$ (triangle, right axis) in function of time.}
\label{fig:NH_cooling}
\end{figure}

 We choose 
100 molecules and a 100 mW power laser propagating along Oz and circular left polarized ($\sigma^+$). The results of the cooling are presented in Figure \ref{fig:NH_cooling}.
We first see that the repumping is efficient and that, despite the transient population of an untrapped potential curve, we do not loose any molecules during the process.
The other important result is that the temperature is reduced, and therefore the phase-space density increases, but that the process stops, and even gets worse after 100 ms.

If, by choosing better laser parameters, the final temperature can be made lower,
we would like to stress that such limited cooling behaviour is quite general and that choosing better cooling strategies or parameters is not so obvious.
Thus, this is an important result appearing in several situations that we have simulated when dealing with zero field trap at the center.
By  looking to the transitions radius and potential energy for the particles, we found that
 with degenerate energy levels near the trap centre (i.e. at low temperature), the spectral or spatial selectivity for the transitions is no longer ensured and that we excite undesirable levels during the process.

\subsection{Cooling in  trap without energy degeneracy}

A simple way to avoid such nasty behaviour is to use a  trap without energy degeneracy, that is with a non zero field at the center,  where the spectral selectivity for the transitions is ensured. Another major advantage of such trap is that the field at the center defines a proper quantification axis than can be used to select even more efficiently the desired transition through proper laser polarizations.

Therefore, in order to lift degeneracies and to facilitate the spectral amplitude selection shaping for the optical pumping we choose here a Ioffe-Pritchard configuration.
 Keeping achievable values \cite{1993PhRvL..70.2257S,1995PhRvA..51...22T,1995PhRvA..52.4004W,2005PhRvL..94g3003G,2006ApPhB..82..533M},
 the magnetic field is taken to be 
$\bm B (\bm R) = B'(X \bm e_{\rm X} - Y \bm e_{\rm Y}) + (B_0 + B'' Z^2 ) \bm e_{\rm Z} $ 
with  $B_0 = 0.1$ T, $B'= 1$ T/cm and  $B'' = 0.1$ T/cm$^2$. 
In order to avoid particles that orbit around the trap center,
we need to
 create asymmetries in the potential. We choose $\alpha = 0.17$ creating an 
 asymmetric configuration with a 1.17 T/cm gradient along X and a 0.83 T/cm gradient along Y.

 We still simulate particles initially at a temperature of $T=100$ mK and randomly  distributed using equipartition of the energy.

Because the trap is shallower, the sample is bigger. The Sisyphus laser has now a waist of 6 cm,
a 5W power and a central frequency such as the resonant transition is for particles having a potential energy of $3 k_B T$ where $T$ is the expected temperature. 
We expect an exponential decay of the temperature $T$, and consequently of the sample size. Thus  the power, the detuning, as well the as the (initially 35 MHz) FWHM Lorentzian laser linewidth (this is to follow the interaction time and required resonance sharpness evolution) of the Sisyphus laser	are decreased exponentially with the same time constant (170 ms).
 The bias field creates a very useful preferential quantization axis (the Z axis) and we thus choose to propagate the lasers along this axis with circular polarization.
For this simulation we use a realistic repumping laser with 10W power, a 5 GHz laser FWHM linewidth in order to repump all needed untrapped particles at a  waist of 3 cm. 
We did not find it useful and necessary to reduce the waist of the lasers and this is a clear simplification for experimentalists.
In order to repump only useful particles, that is the particles having a low enough energy, the spectrum of repumping the laser is flat and cut at a frequency corresponding  to particles in $M''=1$ with a potential energy of $0.4 k_B T$. 
We just mention that spectral shaping for amplitude selection   has experimentally been realized for optical pumping experiments
   \cite{MatthieuViteau07112008,2010NatPh...6..271S,2010NatPh...6..275S,PhysRevLett.109.183001},
  with similar laser power and shaping resolution using  simple broadband diode lasers and diffraction gratings  for the shaping \cite{2009PhRvA..80e1401S,2009NJPh...11e5037S,2010NaPho...4..760C}.

\begin{figure}[ht]
\centering
\resizebox{0.4\textwidth}{!}{
		\rotatebox{-90}{
\includegraphics*[20mm,42mm][187mm,276mm]{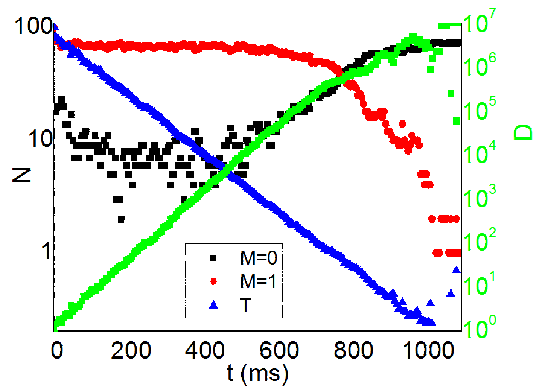}
		}
		}
		\caption{(Color online). Results of NH Sisyphus cooling of molecules in a quadratic (Ioffe-Pritchard) anisotropic  trap. Molecules number $N$ in the $M''=1$ and $M''=0$ states and temperature $T$ are display on the left axes. The relative phase space density $D$ is display in right axes in function of time.}
\label{fig:NH_cooling2}
\end{figure}

The results of the simulation, shown in Fig. \ref{fig:NH_cooling2}, are impressive. With almost no losses, the temperature drops by a factor 1000 and the phase-space density increases by a factor $10^7$ in only one second. Even better results are possible, but we may need to take into account collisional processes for such high phase space densities. We thus believe that further optimization of the parameters is useless. 

	\section{Case with low Franck Condon value: Optical pumping}
	
	With the previous NH case, we have treated a molecule with a very high Franck-Condon factor where vibration is frozen. However several molecules does not have such good properties (see Table \ref{tab:mol}).
	
	For such cases, with bad   Frank-Condon factors, possible improvement of the Sisyphus cooling would consist of  even further reducing the amount of  (absorption)-spontaneous emission cycles,  for instance by using
 an external
cavity	 \cite{2008PhRvA..77b3402L}
or 
 coherent effects where the
spontaneous emission step occurs only after several energy transfer processes 
 \cite{1997PhRvL..78.1420S,2009PhRvA..79f1407H,2010PhRvA..82e3408O,2011PhRvA..84f3401C,2012arXiv1201.1015I}. 
But even, without using such coherent (often comlex) tricks,
	there are at least three possibilities to deal with such difficulties that we shall mention:
:

	\begin{itemize}
		\item Use an optimized single photon cooling to avoid losses due to spontaneous emission.

		We have already studied  this solution and conclude that, if possible experimentally, three times 1D cooling may be more efficient that direct 3D cooling \cite{1995PhRvL..74.2196N,2009PhRvA..80d1401Z}.

		\item Realize a Sisyphus cooling in the  ground state using only ro-vibrational transitions  as done in Ref. \cite{2009PhRvA..80d1401Z,2012Natur.491..570Z}.

		In light diatomic molecules the
 radiative lifetime of the excited ro-vibrational state is not too long and this  solution can be efficiently implement. For instance
the $ v'=1 \rightarrow v''=0$  radiative lifetime of NH(X$^3 \Upsigma^-$) is determined to be 37 ms \cite{2008PhRvL.100h3003C}  (59 ms for OH \cite{2005PhRvL..95a3003V}). Very powerful lasers, such as  single-frequency continuous-wave optical parametric oscillator (OPO),  are nowadays commercially available to drive this transition (at 3.05 $\upmu$m for NH). 
	Due to the long lifetime of the states it is preferable to work only with trappable states. This will often implies to work with $M'=2$ (and thus $J'=2$) upper state because it decays towards $M''=1,2,3$ lower states. Several lasers or microwave sources might be needed to repump these levels.

		\item Use an optical pumping scheme in order to repump the levels.
	\end{itemize}
	
	Obviously we can combined the three possibilities in order to find the best solution to the problem.

	To help for further studies on Sisyphus cooling of such molecules, we deliberately choose to illustrate the third solution  on a system,
  combining several ``bad" points such as: no closed rotational neither vibrational transition scheme and quite poor Franck-Condon factors ($ \leq 0.5$).	 Thus,
we shall  illustrate the method on a textbook example inspired by the CO case, but with scaled internuclear distance to get the Franck Condon factor of 0.5 towards the
    $^3 \Uppi_2$
    lower
  level. Quantities related to this level will be indicated by a double prime $''$. 
	
\begin{figure}[ht]
\centering
\resizebox{0.5\textwidth}{!}{
		\rotatebox{-90}{
\includegraphics*[69mm,42mm][153mm,246mm]{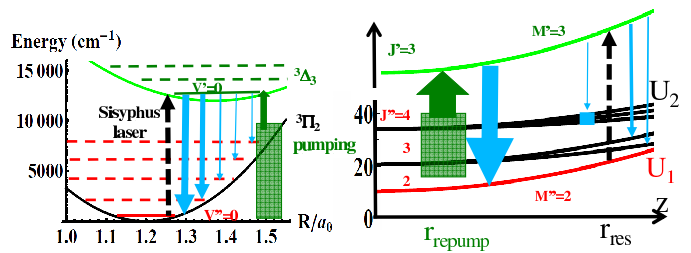}
		}
		}
		\caption{(Color online). Branching ratios,  proportional to the broadening of the blue lines,  between vibrational (left) and rotational (right) levels after
		transfer by the narrowband Sisyphus laser (dashed black) and the spectrally shaped amplitude selection broadband optical pumping (green hatched). $J''=2$ is not pumped for $v''=0$.
	}
\label{fig:FC}
\end{figure}

 Then 
	the  $ ^3 \Updelta_3$  upper level, indicated by a prime $'$,  is preferred to the $ ^3 \Upsigma$  level because it forms with $^3 \Uppi_2$ a closed system for spontaneous emission.
Both potentials  \footnote{We calculate them using 
     CO reduced mass, harmonic potentials and rigid rotor energy levels given by $\hbar \omega (v+1/2) + B J (J+1)$. Equilibrium point, vibrational frequency and rotational constant for respectively lower and higher potential are: $1.2 a_0, 1700$ cm$^{-1}, 1.7$  cm$^{-1}$  and,
     $1.37 a_0, 1200$ cm$^{-1}, 1.3$  cm$^{-1}$. The $v''=0, v'=0$ transition being at 12000 cm$^{-1}$.  
		}  are presented in Fig. \ref{fig:FC} where the Franck-Condon factors, the rotational transitions and the trapping energy states are also displayed. They produce a Franck-Condon value near 0.5 for $v'=0 \leftarrow v''=0$.
 
An obvious choice for the initial trapping $U_1$ is the lowest ro-vibrational level having the strongest trapped state that is the $|v'' J'' M'' \rangle = |0   2 2\rangle''$ level.
By choosing excitation only to 
$|J' M' \rangle = | 3 3\rangle'$, the rotational population can be locked in few trapped levels: $|J'' M'' \rangle \in \{|22,32,33,42,43,44 \rangle'' \}$.
Finally, in order to help  the vibrational cooling, we do not use the
  best existing Franck-Condon factor (on $v'=1 \rightarrow v''=0$)
 but we deal  
  with excitation towards  $v'=0$ to favor decay  to low $v''$ values (see Fig. \ref{fig:FC}). Therefore
for simplicity of both the explanation and the theoretical treatment, all lasers excitations end up to the sole  $|v' J' M' \rangle = |0 3 3\rangle'$. 

   We simulate particles initially at a temperature of 100 mK in an Ioffe trap with
 $B_0 = 0.1$ T,$B'=0.1$ T/cm and  $B'' = 0.1$ T/cm$^2$.

Several laser schemes or simple RF-microwave transfer can be used to make the Sisyphus or pumping transitions. We choose here a simple case with only two lasers: a single one to make the Sisyphus transition and a second one as the pumping laser near the trap center which shall also make optical pumping to bring back population from others ro-vibrational states. Thus,
	close to the trap center the (re-)pumping laser should bring the molecules back into the $ |0  2 2\rangle''$ state.  A single laser would
 create 
 a cylinder pumping zone. This would be
a problem for particles moving mostly along its axis because they will never leave the cylinder and will be always under the effect of the repumping light. To be closer to the ideal situation of Fig.  \ref{fig:opt_pump}, with small $r_{\rm repump}$, 
we thus create a more spherical radial zone by dividing the  pumping laser in three beams	propagating along Ox, Oy and Oz. They are focused at the trap center  on an initial 4 mm waist, all circularly left
 polarized (corresponding to $\upsigma_+$ polarized for a quantization axis  along the beam propagation) and
	 their spectrum, 60 mW/cm$^{-1}$ uniform power density, are shaped (amplitude selection)  to drive only transitions towards $|0 3 3\rangle'$.
  Finally, in order to accumulate population into the  $| 0 2 2 \rangle''$ level,  all transitions from this state are removed by the spectral shaping.

Several possible experimental realizations exist for the repumping laser: spanning from femtosecond lasers, diode lasers, LED light, or even a supercontinuum laser, nowadays  spanning 0.4-2.3 $\mu$m  with 50 mW/nm uniform power density \cite{2010JOpt...12k3001T}. We suggest a simple experimental setup based on the use of several individual lasers driving individually the vibrational transitions. This would requires five diode lasers to cover the $v''=0-4 \rightarrow v'=0$ transitions respectively at $833, 971, 1163, 1449$ and $1923$ nm wavelength, that are all available commercially.
The presence of the $B_0 = 0.1$ T  bias field,  lifts the energy levels through the Zeeman effect by $ 0.05\, $cm$^{-1}$. The required GHz range resolution is at the edge of what can be achieved by using a simple grating configuration, but higher resolution optical pulse shaping methods can be used \cite{2011OptCo.284.3669W}. We mention the optical arbitrary waveform  line-by-line
pulse shaping technique \cite{2010NaPho...4..760C}, based, for instance, on arrayed waveguide routers
or
 liquid
crystals on silicon 
 \cite{Wilson:07,Frumker:07,Fontaine:08}, or the virtually imaged phased arrays that combine the
high spectral resolution potential of etalons with the spectral
disperser functionality of gratings \cite{Shirasaki:96}. Such devices have been applied to realize
pulse shapers with sub-GHz spectral resolution \cite{Xiao:05,Willits:12,Supradeepa:08}. But one simple method is the scanning diode method used in my group \cite{PhysRevLett.109.183001}.

	The Sisyphus transition, $|0   2 2\rangle'' \rightarrow |0 3 3\rangle'$, is performed using a  1 W laser at 833 nm, 
	10 MHz linewidth,
	covering the whole sample with its
	2 cm waist, left circularly 
 polarized, propagating along Oz, with initial detuning chosen to initially transfer even very energetic particles with potential energy of $3.5 k_B 100$ mK. 

The waist size of repumping ans Sisyphus lasers, as well as their powers and the detunings	are decreased exponentially with a time constant of 500 ms.

\begin{figure}[ht]
\centering
\resizebox{0.5\textwidth}{!}{
		\rotatebox{-90}{
\includegraphics*[16mm,32mm][172mm,274mm]{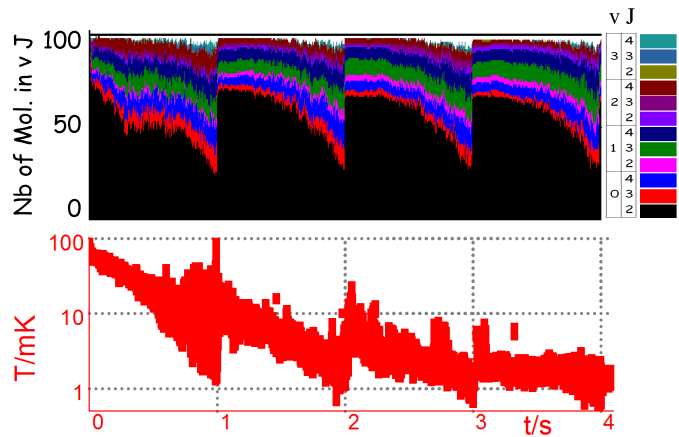}
		}
		}
		\caption{(Color online). Time evolution of the (stacked) number $N$ of
	 particles in $|v'' J''\rangle$ state 
		and the (three axis) averaged
		temperature $T$ of the particles in $|022\rangle''$. The  sequence is repeated every 1 s.
	}
\label{fig:results}
\end{figure}

  The results of the simulation is presented in Fig. \ref{fig:results} where the temperatures are calculated from the  standard deviation of the histogrammed velocities.
 25\% of the $|022\rangle''$ molecules are cooled from 100 mK to  1 mK within 1 s.
Other molecular states are simply not repumped because they are in larger orbits, but no molecules are lost. Thus,
repeating the same process every 1 s greatly improves  the results with 60\% of the molecules with temperature below 1 mK after 4 s.

Clearly, there is plenty of room for  optimization \footnote{In our simple scheme a lot of particles are transferred into $|v''\neq 0, J''=2, M''=2 \rangle$ levels, having same trapping potential	than the initial one and not a less stiff one as desired for good Sisyphus cooling. This is a waste of time because a pumping step is required before another cycle.} and
 the lasers wavelengths, sizes, locations, polarizations and chosen transitions can be further optimized.  For instance, simulation of a perfect Sisyphus scheme where particles are transferred and repumped when reaching their higher and lower radial coordinates
 leads to
 100\% of the molecules being cooled, from 100mK down to 1mK, within only 0.15 s.


	\section{Generalization of the method}
	
\begin{figure}[ht]
\centering
\resizebox{0.5\textwidth}{!}{
		\rotatebox{-90}{
\includegraphics*[10mm,23mm][167mm,267mm]{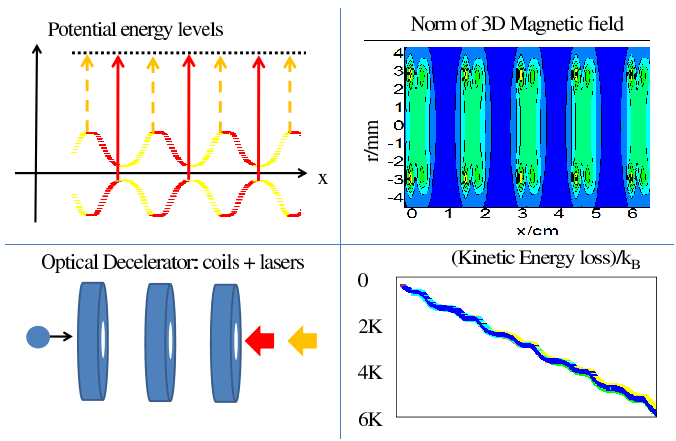}
		}
		}
		\caption{(Color online). Illustration of optical deceleration of a molecular beam using Sisyphus cooling when molecules travels through magnetic coils. Left Panel: illustration of the principle. Right Panel: Simulation using pairs of 3mm radius Helmholtz coils separated by 1.5cm and creating a 1T axial field. The circularly polarized lasers are 5 W focused with 1.5 mm waist, 1 GHz laser linewidth. The monokinetic molecular beams cover 1 mm radially.  
	}
\label{Fig:decel}
\end{figure}

Before to conclude we would like to mention that
the Sisyphus cooling method can be generalized with more complex potentials or absorption schemes for instance by during decelerators stages.
 We have illustrated in Fig. \ref{Fig:decel}
	such idea    with molecules traveling repeatedly through uphill  potentials. The idea is similar to the one proposed in Ref. \cite{2009PhRvA..79f1407H} but now with a spontaneous emission step. The 3D simulation is done using a simple ``toy model"  spin $1/2$ molecule (four levels A $^2 \Uppi_{1/2} M'_J=±1/2 \leftarrow {\rm X} ^2 \Upsigma, M''_J=±1/2$	transition) but is obviously more general. 
	The global dynamics of the decelerator looks promising with negligible effect due to the
transverse motion and only small Kinetic Energy dispersion  due to photon recoil energy transfer.
	This scheme can be realized using permanent magnet and would allow continuous cooling.

\section{Conclusion}

In conclusion, the fact that
Sisyphus and optical pumping methods  requires only a small number of spontaneous emission steps makes it especially useful for cold molecular systems with internal state modifications during the individual process. Similarly,
 an optical one-way pumping method can also help to realize transfer to a trap after a deceleration or a buffer gas cooling stage \cite{2008PhRvL.100x0407T,2011PhRvL.106p3002F,2009NJPh...11e5046N,2010PhRvA..82b3419S}. 
We stress here that a full and detailed simulation of the laser interaction is very important, a less detailed simulation can lead to incorrect results. We found that the laser power and its linewidth are parameters that are not so simple to tune. A relatively fast particle, a low power laser, an incorrect local polarization or a too small laser linewidth can result in transitions that are not effective. On the contrary a slow particle, a  high power laser or  a large linewidth can result in transitions occurring too early, out of resonance and not at the desired location nor energy location. We also found that a small modification of the trap can totally modify the efficiency of the cooling due to possible angular momentum quasi-conservation.
The spectral or spatial selectivity for the transitions have to be ensured and  care has to be taken not to excite undesirable levels. 
A trap which lifts energy degeneracy allows simpler spectral selection of the transitions and in this case
the cooling is found to be much easier by using preferential quantification axis allowing to control transition using light polarization.

We also show that three times 1D cooling may be simpler and more efficient that direct 3D cooling \cite{1995PhRvL..74.2196N,2009PhRvA..80d1401Z}. 
Indeed, in 1D
the quantification axis is always the same and transitions can thus be controlled using light polarization, which can be extremely useful to excite only state which are trapped (for instance by controlling the Zeeman sub-level in a magnetic trap).
Furthermore, rotation of phase space may be also advantageously used with
only one laser direction to cool the 3D sample.
 
Obviously the Sisyphus method   
is very easy to implement for trapped
   quasi closed systems   \cite{2004EPJD...31..395D,2009PhRvL.103v3001S,2010PhRvA..82e2521I,2011PhRvA..83e3404N} and can then be a very impressive complement to standard laser cooling. For trapped species,
 the photon transition rate can  be very slow
and the scheme is particularly suitable for particles that require deep-UV lasers for electronic excitations. It also allows for the use of pure ro-vibrational transitions or of quasi-forbidden electronic transitions.
Using  state of the art (GHz) shaping resolution \cite{2010NaPho...4..760C,2011OptCo.284.3669W,Willits:12,PhysRevLett.109.183001} the optical pumping step should not be a problem. 

The amount of possible cooling schemes, using
optical, magnetic, electric or gravitational forces combined with
laser (or Radio-Frequency) transitions of any kind (Raman, coherent, stimulated, pulsed, shaped, multi-photonics, black-body ...) is so large and the method so versatile that it can be implemented to several  species including ions \cite{2011NJPh...13f3023N,2011PhRvA..83e3404N}, ranging from simple atoms  to  polyatomic molecules, that are difficult to laser cool, such as (anti-)hydrogen. It can be generalized to realize 
continuous cooling in new type of Stark or Zeeman decelerators.
  The low temperature achieved can then be a new starting point for further studies  in precision measurements or in chemical control of  highly correlated systems when collisions between molecules start to play a bigger role for lower
temperatures.

Aknowlegments:
I am indebted to N. Vanhaecke and H. Lignier for useful discussions. 
The research leading to these results has received funding from the European
Research Council under the grant agreement n.~277762 COLDNANO.

\appendix

\section{Optimized strategy: one-way single photon cooling VS Sisyphus cooling}
\label{cool:strategy}

\subsection{Trajectories in 3D  traps}

We describe (for simplicity) the trajectories in the case of harmonic trap but the discussion is general.

Trajectories in 3D harmonic traps are given, with  $(x,y,z)=(r_1,r_2,r_3) $ notation, by
$U_1(\bm r(t)) =\frac{1}{2} m \sum_i \omega_i^2 r_i(t)^2 $ with
$r_i(t) = r_i^0 \cos(\phi_i (t))$ and  $\phi_i (t) = \phi_i^0 + \omega_i t$.
The total energy $E =\frac{1}{2} m \sum_i \omega_i^2 {r_i^0}^2 $ is a conserved quantity. We want to reduce the kinetic energy
$E_{\rm kin}(t) = E- U_1(\bm r(t))$.
This  constrains the  trap geometry such that it should avoid angular momentum conservation
 where some  
particles can orbit around the trap center  with a kinetic energy that never reduces: for instance if $\omega_i = \omega_j$ and  $\phi_i^0=\phi_j^0 + \uppi/2$ (dashed trajectory in Fig. \ref{fig_remov_energy}).
As studied in Ref. \cite{2010PhRvA..82c3415C}, we should look for a ``chaotic" regime, with
various different trajectories, increasing
the likelihood of meeting the best condition for single photon cooling where  kinetic energy is near zero: $U_1(\bm r(t)) \approx E$ that is when $r_1,r_2,r_3$ are all extremum. 
In the harmonic case, the coordinate $r_1$ becomes positive maximum when $\phi_1(t) = 0 [2 \uppi]$, at times $t_{k_1} = t_0 + k_1 T_1 $, for $k_1=0,1,\cdots,N_0$,  spaced by the oscillation time  $T_1 = 2 \uppi /\omega_1$. 
During these $N_0$ oscillations we want to have, at $t_{k_2}$,  $r_2 (t_{k_2}) =  r_2^0 \cos( \phi_2^0 + \omega_2 t_0  + 2 \uppi k_2 \omega_2/\omega_1 )  $ almost maximum. This optimal condition ($r_i$ maximum) should occur for all particles and thus for any trajectory independently of the initial condition $\phi_2^0$. A simple choice to realize such situation is $ \omega_2/\omega_1 = 1/N_0 $   because during the $N_0$ periods, the phase $\phi_2(t_{k_1})$ spans the space $[0,2\uppi]$ with interval of $2\uppi/N_0$, and thus we found one $\phi_2 (t_{k_2}) [2 \uppi] $ which approaches $0$  with a maximum error of $\Delta \phi =  \uppi /N_0$.  This is the best possible result with $N_0$ points. 
 Similarly, in $N_0^2$ oscillations, the choice $ \omega_3/\omega_2 = 1/N_0 $  will allow $\phi_3 (t_{k_2})$ to spans the space $[0,2\uppi]$ with interval of $2\uppi/N_0$. At a time $t_{k_3}$ where all positions are almost maximum, the maximum kinetic energy is 
$E- U_1(\bm r(t_{k_3})) = \frac{1}{2} m \sum_i  \omega_i^2 (r_i^0)^2 \sin^2(\phi_i(t_{k_3})) \approx (\Delta \phi)^2 E $.

 From this study we see that ratio of frequencies on the order of $1/N_0$ are good.  Coprime or incommensurable ratios may be also interesting but a simple (from the experimental point of view) choice is to modify slightly the trap geometry by a ratio
$ 1-1/N_0 $.
We have tested this hypothesis in the case of a simple quadrupole magnetic trap where the magnetic field is
$\bm B (\bm R) = B'_0(1+\alpha)  X \bm e_{\rm X} + B'_0(1-\alpha)  Y \bm e_{\rm Y}) - 2 B'_0 Z \bm e_{\rm Z} $ 
with  $B'_0=100 $ T/m and $\alpha $ is a parameter that controls the anisotropicity of the trap. 
We have looked for both for linear traps, i.e. with a state with a linear Zeeman effect, and for quadratic traps, i.e. with a state with a quadratic Zeeman effect: to be more precise
the molecules used is NH in the exact same configuration than used for Fig. \ref{fig:NH_X}: using  $J''=1,M''=1$ (linear)state or the $J''=1,M''=0$ (quadratic state).  The typical "oscillation time" $T_1$ is 2 ms.

\begin{figure}[ht]
\centering
\resizebox{0.5\textwidth}{!}{
		\rotatebox{-90}{
\includegraphics*[10mm,22mm][201mm,160mm]{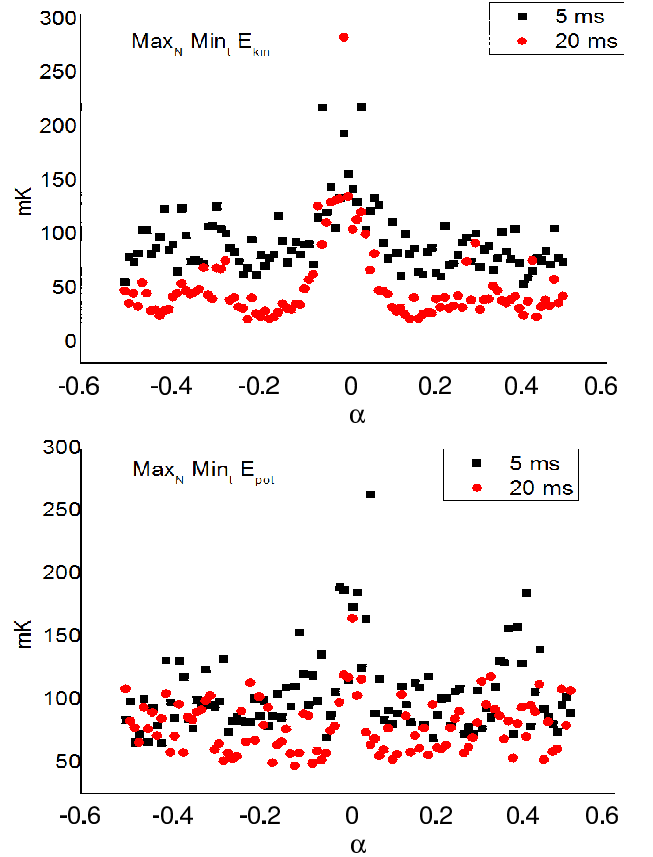}
		}
		}
	\caption{(Color online). Effect of the  trap anisotropy,  on the minimum kinetic $E_{\rm kin}$ (lower panel for a quadratic trap) and  potential energy  $E_{\rm pot}$ (higher panel for a linear trap) reached after 5 ms (black square) or 20 ms (red circle) evolution time of $N=1000$ molecules initially at 100 mK. The worst energy case among the $N=1000$ molecules is plotted, in temperature units, in function of $\alpha$. $1-\alpha$ can be seen as the aspect ratio of the trap (see text for detail).}
	\label{fig:aniso_trap}
\end{figure}

 In order to see if the geometry allows trajectories to reach the best condition for cooling, i.e. with  kinetic energy near zero for the Sisyphus pumping step and with potential energy near zero for the repumping step, we use 1000 molecules initially in thermal equilibrium. For each molecule we look for the minimal kinetic and potential energy reached during a given time: 5 ms  or 20 ms, the first case corresponding to $N_0^2 \sim 5/2$ and the second to $N_0^2 \sim 20/2$. Then we took the worst molecules and plotted the result in Fig. \ref{fig:aniso_trap} in function of the anisotropic parameter $\alpha$. 
The simulation confirms that the anisotropicity of the trap is a crucial parameter and that $\alpha \sim 1/N_0$ and a minimal energy of $E/N_0^2$ are correct assumptions. In the case of an $X,Y$ symetric trap ($\alpha=0$) it is clear than some molecules do not transfer   their energy between kinetic and potential.
Clearly the longer time we allow for the molecule to move the better the results is.

\subsection{Energy or spatial selection  in 3D  traps}

With this simulation and
without  entering into a more detail study \footnote{A proper study of a general, non harmonic, trap case should be done using revolving orbits, orbital precession,  recurrence or Poincaré analysis for the investigation of this dynamical systems. 
This should illustrate in more detail which trap geometry is an efficient one.}, our
 general conclusion is
 that, in $N_0^2$ times the typical oscillation time $T_1$ along the fastest axis (say $x$), a good trap geometry, should allow the trajectories to reach the maximum energy $E$ within $\Delta E \approx E/N_0^2$  \footnote{By removing the positivity condition on the coordinates, in the previous discussion, we have twice more maximum and $\Delta \phi$ becomes  $  \uppi /2 N_0 \sim 1/N_0$.}. 

\begin{figure}[ht]
\resizebox{0.5\textwidth}{!}{
		\rotatebox{-90}{
\includegraphics*[9mm,22mm][118mm,228mm]{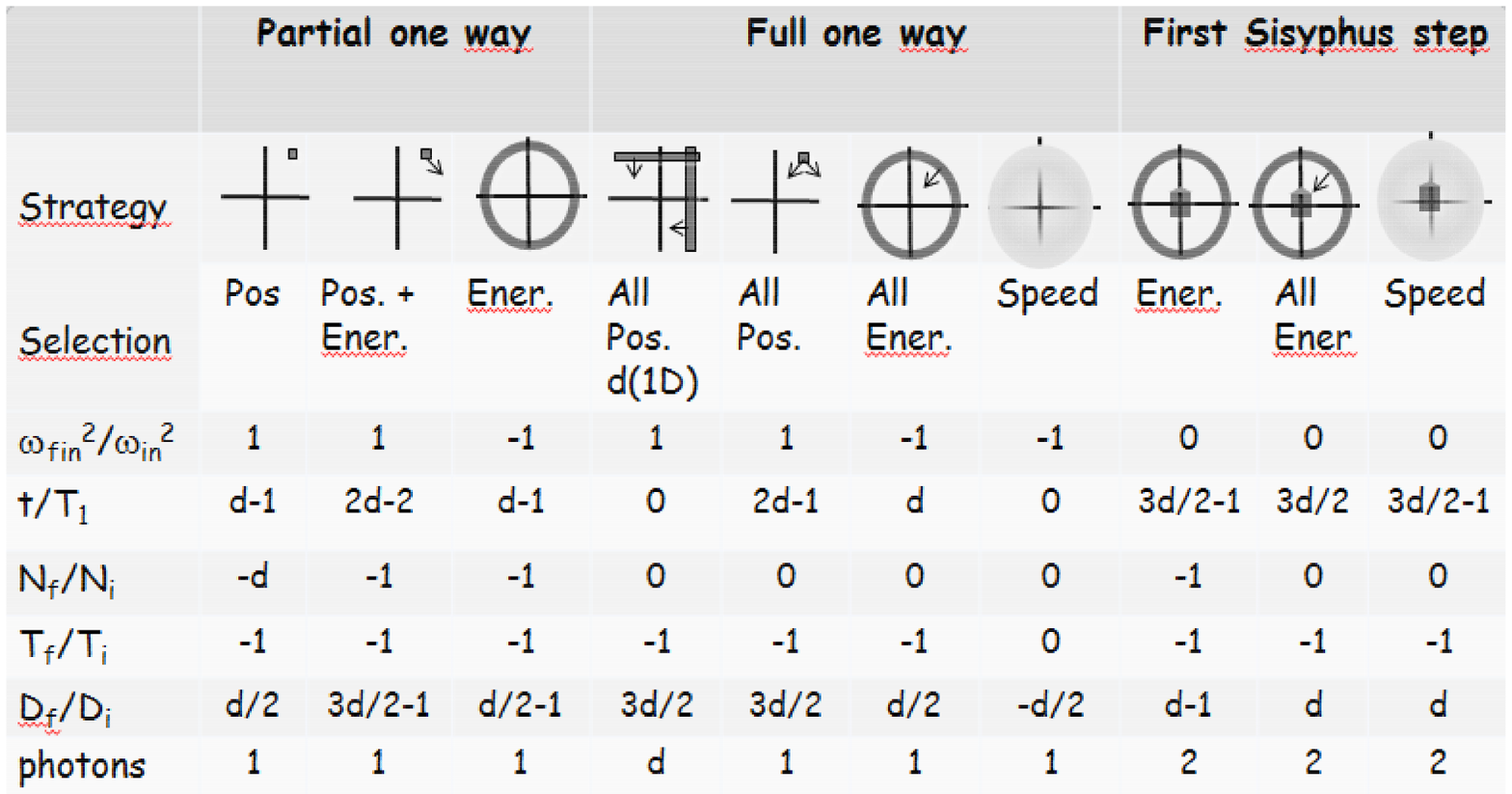}
		}
		}
	\caption{(Color online). Typical evolution of number of molecules $N$, Temperature $T$ (or energy $E$), phase space density $D$ and time required $t$, for the sample under different 
	$d$ dimensional cooling strategies  (pos. means a selection in position as in  Fig. \ref{fig_remov_energy} 1), Ener. a selection in energy as in  Fig. \ref{fig_remov_energy} 2) and Speed a selection with rate proportional to velocity as in  Fig. \ref{fig_remov_energy} 3). The values are given in the  harmonic trap cases with angular frequencies initial $\omega_{\rm in}$ and final $\omega_{\rm fin}$. The  {\bf partial one-way} ones takes only part of the population (with a box of given energy or spatial range). The {\bf Full one-way} strategies move the previous box to catch all particles. The {\bf first Sisyphe step} (step 3 of Fig. \ref{fig:opt_pump}) put the population back in the original potential. The   {\bf Sisyphus} is thus simply the repetition of this step.  The parameters, except the photons numbers, are given in power of $N_0=k_{\rm B} T/\Delta E$ which is a parameter indicating the number of slices  of the sample	
 over which we catch particle by spacial (or energy) steps. The time evolution is given, in the "t" line, in function of the typical oscillation time $T_1$ along the fastest axis motion: $d-1$ indicates for instance that the time evolve like $t\propto T_1 N_0^{d-1}$.
}
	\label{tab:scaling}
\end{figure}

Finally, to be more precise we also have to take into account that the Sisyphus transition should occur  when the particle stays in the $E \pm \Delta E$ energy range, that is during a time $\Delta t \approx T_1/N_0$. 
This is $N_0^3$ times smaller than the overall time motion and, due to interaction  time or power broadening of the transition, this can lead to
 lower energy resolution than the expected $\Delta E \approx E/N_0^2$ . The study of this effect is in fact the core of the simulation and will be done later numerically. However, 
a simple  Lorentzian shape for  the excitation rate (cf Eq. \ref{eq_rate}) leads to an energy resolution $N_0^{3/2}$ worse
and thus  $\Delta E \approx E/N_0^{1/2}$. 
In conclusion, in order to have simple calculation, we assume 
an intermediate case of $\Delta E \approx E/N_0$ for all cooling strategies that we study.
 We have sketched them and the results in the Table \ref{tab:scaling}.

The third column of Table  \ref{tab:scaling} deals with
the energy removal strategy (case 2 of Fig.  \ref{fig_remov_energy}).
By looking at the typical case of $E \sim k_B T$, and with $\delta E \approx E/N_0 $, we see that 
  a single photon can lead to a reduction of the energy, or the temperature, by a factor $N_0$.  
 The  depth (and shape) of the capture potential $U_2$ can thus be optimized, to be on the order of $k_B T/N_0$, i.e. in the harmonic case with frequencies ${\omega'}_i^2 \sim \omega_i^2/N_0$.  From an experimental point of view, such a flat potential
 can be realized using combined Zeeman and Stark effects, for instance \cite{2012PhRvA..85e3406S}. In this case the density is unchanged and the phase space density $D= N \left( \frac{\hbar \omega}{k_B T} \right)^3$ increases by a factor $ N_0^{1/2}$ in the 3D case and $ N_0^{d/2 -1}$ in the $d$ dimensional case.

It is possible to increase furthermore the phase space density by transferring the particles in a tighter trap without heating. This can be done by
  catching them in a tight trap, cf black box in Fig.  \ref{fig_remov_energy} Low Left Panel 1. This is the case study in the first column of table \ref{tab:scaling}. 
The best spatial catch is when the particles reach  the transfer zone of $ \Delta E \approx E/N_0$. 
However,  the spatially located box is able to access only the turning points concentrated within its small region. For instance, from the figure Fig.  \ref{fig_remov_energy}, we clearly see that if the box was placed elsewhere with the same energy, the trajectory  would never cross it. 
If we want to catch all particles, we have either to move the box to cover the all spatial transfer zone of $ \Delta E \approx E/N_0$ (cf second and fifth column of table \ref{tab:scaling}), or to move the trajectories to the box (using a small anharmonicity).
Here again in a 3D case we can cover this space in $N_0^2$ spots.
	
Another strategy is to  use a second photon 
  to pump the molecules towards a tighter trap when the molecules are close to the trap center, using for instance the Sisyphus cooling process (step 3 of Fig. \ref{fig:opt_pump}, and first column of ``Sisyphus step" of table \ref{tab:scaling}). 
In this case,
 very similar to the dimple trap trick used in evaporative cooling \cite{1998PhRvL..81.2194S,2006PhRvA..73d3410C}, we need to transfer the molecule when the position is as close a possible to the center i.e.  where the kinetic energy is maximal. The exact same discussion indicates that 
in $N_0^2$ times the typical oscillation time $T'_1 \sim \sqrt{N_0} T_1$ in the new trap,
 the trajectories reach the maximal kinetic energy $E' =  E/N_0$ within $\Delta E' \approx E'/N_0$.

The values given here are only indicative. But they can help to choose the parameters. For instance with $T_1 \sim 10$ ms ($\omega_1\sim 2 \uppi (100 $ Hz)), a value of $N_0 = 4 $  is possible even for a Full one way strategy spanning the all positions. Indeed, this gives in  3D ($d=3)$ a cooling time $N_0^5 T_1$ of 10 s which may be feasible. This should allow a huge gain in phase space density of $4^{3d/2} \sim 500$
The first (1D) Sisyphus cooling scheme, leading to Eq. \ref{eq_T_1D},  was done with $N_0=2$, i.e. $\omega' = \omega/2$.

\section{Rate equations, Kinetic Monte Carlo and N-body integrator used}
\label{appendix}

\subsection{Polarization and Rotation matrices}


 When dealing with rotation, and angular momentum, several convention exist for the phase relation between coefficients, for the sens of rotations or for the name of the Euler angles. A good summary of the existing convention is given by \cite{varshalovich1987quantum}. 
We deal here with several different frames or axis: the lab fixed one (given by the vacuum chamber), the laser axis (given by the wavevector $\bm k$);  for a given molecule we have the local field $\bm B$ or $\bm E$ axis (depending on the position of the particle) and the molecular axis.

We use capital letter X,Y,Z for space fixed frame, with direction vectors $( \bf{e}_{\rm X}, \bf{e}_{\rm Y}, \bf{e}_{\rm Z})$. We use lower case letter for the
 local frame $( \bf{e}_{\rm x}, \bf{e}_{\rm y}, \bf{e}_{\rm z})$
with $\bm B$ (or $\bm E$) field is along the Oz. 
 We shall also use the (covariant) spherical and helicity basis vectors
$ {\bf e}_0 = {\bf e}_z, {\bf e}_{\pm 1} =  \mp \frac{{\bf e}_x \pm
i {\bf e}_y} { \sqrt{2}}$ because they directly 
express light polarization. Another frame
$( \bf{e}'_{\rm x}, \bf{e}'_{\rm y}, \bf{e}'_{\rm z} = \bm{k}/\| \bm{k} \|)$ is linked with the laser propagation. The original laser polarization is described in the helicity basis
$ {\bf e}'_0 = {\bf e}'_z, {\bf e}'_{\pm 1} =  \mp \frac{{\bf e}'_x \pm
i {\bf e}'_y} { \sqrt{2}}$.
We
can write the laser polarization vector in the two frames  as \cite{varshalovich1987quantum} Eq. 1-(49):
 $ {\bm \epsilon} =
	 \sum_{p'= -1,0,+1} {\epsilon'}^\ast_{p'} 
{\bf e'}_{p'} =  \sum_{p= -1,0,+1} \epsilon_p^\ast 
{\bf e}_{p} $, with
$ \epsilon_{p} = 
\sum_{p' =0,\pm 1 } (-1)^{p'} \epsilon'_{p'} 
D_{p p'}^{\ast (1)} (0,\theta,\varphi)$. We find:
\begin{eqnarray*}
\epsilon_{+1}   & = & \frac{1+\cos \theta}{2} {\rm e}^{- i \varphi} \epsilon'_{+1}  + \frac{\sin \theta}{\sqrt{2}}  \epsilon'_{0} + \frac{1-\cos \theta}{2} {\rm e}^{ i \varphi} \epsilon'_{-1}  \\
\epsilon_{0}   & = & - \frac{\sin \theta}{\sqrt{2}} {\rm e}^{- i \varphi} \epsilon'_{+1}  + \cos \theta  \epsilon'_{0} + \frac{\sin \theta}{\sqrt{2}} {\rm e}^{i \varphi} \epsilon'_{-1} \\
\epsilon_{-1}   & = & \frac{1-\cos \theta}{2} {\rm e}^{- i \varphi} \epsilon'_{+1}  - \frac{\sin \theta}{\sqrt{2}}  \epsilon'_{0} + \frac{1+\cos \theta}{2} {\rm e}^{ i \varphi} \epsilon'_{-1}
\end{eqnarray*}
where $\theta = (\widehat{Oz,Oz'}) = (\widehat{{\bm e}_z,{\bm k}})$ is the angle between the  local field and the laser propagation axis.
When dealing with pure (circular)  polarization and because the transition probability depend only on the modulus square of the transition amplitude the value of $\phi$ becomes irrelevant and we choose $\phi =0$.

\subsection{Transition Matrix Elements and Line Strengths}

The electromagnetic field, due to the lasers $\rm L$, can be written
$${\bm E}(\bm r,t) = \frac{1}{2}\sum_{\rm L}    \left[ {\bm E}_{\rm L} e^{i(  {\bm k}_{\rm L}. {\bm r} - \Phi_{\rm L}
(t))} +  {\bm E}_{\rm L}^\ast e^{-i(  {\bm k}_{\rm L}. {\bm r} - \Phi_{\rm L}
(t))}  \right] $$
 For each laser $\rm L$
	the polarization vector, defined by 
	$ {\bm E}_{\rm L}(\bm r,t) = E_{\rm L}(\bm r,t) {\bm \epsilon}_{\rm L} $, is (written directly in the local axis) $ {\bm \epsilon}_{\rm L} =
	 \sum_{p= -1,0,+1} \epsilon_p^\ast 
{\bf e}_{p}  $ and 
 the irradiance, called improperly  intensity, is 
\begin{equation}
I_{\rm L}= \varepsilon_0 |E_{\rm L}|^2 c/2 \label{eq:intensity}
\end{equation}
The light-molecule interaction hamiltonian is $- {\hat  {\bm \mu}}.{\bm E}$, where $\hat  {\bm \mu}$ is
the electric dipole moment operator.

	The Doppler effect, and the laser linewidth, are taken into account by writing $\Phi_{\rm L} (t) = (\omega_{\rm L}  - {\bm k}_{\rm L}. {\bm v}) t + \Phi^{\rm L} (t)$. Where $\Phi^{\rm L} (t) $ is  a fluctuating phase. As shown below, or through the Wiener-Khinchin-(Einstein-Kolmogorov)  theorem,  its statistical average is linked to  the
spectral irradiance distribution $I_{\upomega} (\omega)$. In the simple Lorenzian case
$I_{\upomega} (\omega)
 = \frac{I_{\rm L}}{ \uppi} \frac{\Gamma_{\rm L}/2}{(\omega - \omega_{\rm L})^2 + (\Gamma_{\rm L}/2)^2} $ with FWHM
$\Gamma_{\rm L}$.

We shall only describe the notation for Hund's case a, but 
for more complex cases I use the known molecular parameter of the molecule and then the transition strength and Zeeman or Stark effect shifts
 ($\Delta E = E_0 + {\rm Sign}(C)[-\Delta/2 + \sqrt{\Delta^2/4 + C^2 F^2}]$ for the of each levels where $F$ is the field)
 given by PGOPHER (Program for Simulating Rotational Structure) see also \cite{BrownCarrington2003}.
We  use the simplified notation $ \langle ' | {\hat  {\bm \mu}} .\bm \epsilon^{\ast} | ''
\rangle =  \langle \Upomega' v' J' M' | {\hat  {\bm \mu}} .\bm \epsilon^{\ast} | \Upomega'' v'' J'' M''
\rangle$ where the electronical part is simply noted $| \Upomega \rangle$, the vibrational $| v \rangle$ and the rotational $ | J M \Upomega \rangle$.
The integration on the electronical coordinates  $\bm r_i$ gives first the electronic dipole moment
$\mu^{\rm el}(R) =
\left\langle  \Upomega' \left|  
q_e \sum_{i=1} {\bm r}_i.{\bf e}_q^\ast \right| 
\Upomega''
\right\rangle (R)$ depending on the internuclear distance $R$.
With the short notation that $q$ refers  to  vector quantized on the molecular frame whereas $p$ refer to vector on the local (field)  frame.
The integration on the vibrational coordinate $R$ gives the dipole moment of the transition
$ 
\mu_q^{v' v''} = \left\langle v' \left|
\mu^{\rm el}(R) \right| v'' \right\rangle$. If $\mu^{\rm el} (R)$ is not constant we can assume a linear variation in $R$ which leads to
$$\mu_q^{v' v''} = \langle v' | v'' \rangle   \mu^{\rm el} (R_{v' v''})  $$
 where the $R_{v' v''} = \langle v' | R | v'' \rangle$ centroid is the ``distance" where the transition takes (classically speaking) place.
When no other information is available  we take the dipole moment of the transition from  Einstein's formula for the lifetime $\tau'$ of the upper state $\tau'^{-1} \approx \frac{ \omega_{\Upomega' \Upomega''}^3}{3\upepsilon_0 c^3 \hbar \uppi}  \left( \mu^{\rm el} \right)^2$.

The integration over the rotational coordinates leads to the final results for the  Rabi Frequency $\Omega$ of 
 the transition:
\begin{eqnarray}
\hbar \Omega &=&  \langle ' | { \hat {\bm  \mu}}.{\bm E}_{\rm L} | '' \rangle =  E_{\rm L} \mu \epsilon_p^*  \label{eq:rabi} \\
& = & E_L \epsilon_p^*  \langle v' | v'' \rangle   \mu^{\rm el} (R_{v' v''})  \times \nonumber \\ 
&  &  \langle J' \Upomega' , 1 q | J'' \Upomega'' \rangle \langle J' M' , 1 -p | J'' M'' \rangle \nonumber
\end{eqnarray}
 where
 $p = M'' - M'$ and $q = \Upomega'' - \Upomega'$ are the only values leading to non zero amplitude.

The intensity of a transition is thus proportional to the Franck-Condon factor $|\langle v' | v'' \rangle|^2$.

\subsection{Laser interaction}

\begin{figure}[ht]
\centering
\resizebox{0.4\textwidth}{!}{
		\rotatebox{-90}{
\includegraphics*[35mm,83mm][150mm,190mm]{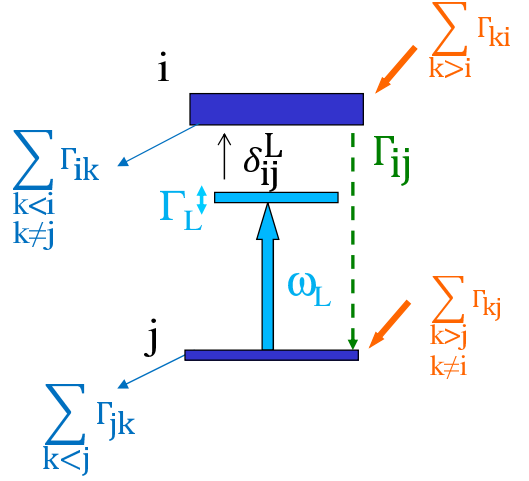}
		}
		}
		\caption{(Color online). Schematics of a levels i and j interacting with a broadband laser and with other levels.}
\label{fig:two_level}
\end{figure}
The optical Bloch equations for the density matrix element $\rho_{\rm i j} = \langle {\rm i} | \hat \rho | {\rm j} \rangle = \rho_{\rm j i}^\ast$  are \cite{CDG2}:
\begin{eqnarray*}
	  	\dot	 \rho_{\rm i i}  & = & \sum_{\rm j} \Gamma_{\rm j i}  \rho_{\rm j j}  
			- \sum_{\rm j} \Gamma_{\rm i j}  \rho_{\rm i i} + 
	\frac{i}{\hbar} \sum_{\rm j} \Big[
 { \hat {\bm  \mu}}_{\rm i j}.{\bm E}   \rho_{\rm j i}  -  \rho_{\rm i j} { \hat {\bm  \mu}}_{\rm j i}.{\bm E} \Big] \\
\dot	 \rho_{\rm i  \neq j}  & = &  -i \omega_{\rm i j} \rho_{\rm i j} -  \frac{ \Gamma_{\rm coll} + \sum_{\rm k} (\Gamma_{\rm i k} + \Gamma_{\rm j k} )
	 }{2} \rho_{\rm i j} + \\
	& & \frac{i}{\hbar} \sum_{\rm k} \Big[
 { \hat {\bm  \mu}}_{\rm i k}.{\bm E}   \rho_{\rm k j}  -  \rho_{\rm i k} { \hat {\bm  \mu}}_{\rm k j}.{\bm E} \Big] 
	\end{eqnarray*}
	Where for completeness we have added some possible collisional dephazing through $\Gamma_{\rm coll} $.
The rate equation we want to derive can be justified only using  several approximations
 \cite{CDG2,2012AmJPh..80..882H} ; especially 
when coherence  (non diagonal) terms are small and thus a dominant part is played by the population (diagonal) terms.
We thus keep only the terms $  \rho_{\rm k j}$  or $ \rho_{\rm i k}$ when $\rm k=i$ or $\rm k=j$, even if this prevent to simulate some dark states coherent schemes \footnote{It is sometimes possible to use the rate equations, even with coherent dark states, simply by changing from the initial basis states to a new basis which includes the bright and dark states such as $(|i\rangle \pm |j\rangle)/\sqrt{2}$.}

Under this approximation we have:
\begin{eqnarray*}
	  	\dot	 \rho_{\rm i i}  & = &  \sum_{\rm j} \Gamma_{\rm j i}  \rho_{\rm j j}  
			- \sum_{\rm j} \Gamma_{\rm i j}  \rho_{\rm i i}  
			- \sum_{\rm j} {\cal I}m \left(\tilde \rho_{\rm j i} \sum_{\rm L} {\Omega_{  \rm i j}^{\rm L}} 	  e^{ i( \Phi_{\rm i j } - \Phi_{\rm L} ) }\right)  \\
	\dot{	  \tilde \rho}_{\rm i j}  & = &  - i \left(\omega_{\rm i j}
	- {\rm d} \Phi_{\rm { i j}}/ {\rm d} t  \right)
	{\tilde \rho}_{\rm i j} -\frac{\Gamma_{\rm coll}+\sum_{\rm k} (\Gamma_{\rm i k} + \Gamma_{\rm j k} )
	}{2}   		
{\tilde \rho}_{\rm i j}  +  \\
 & &
	\frac{i}{2}  ( \rho_{\rm j j} -  \rho_{\rm i i}  ) \sum_{\rm L'} \left[
 {\Omega_{  \rm i j}^{\rm L'}} e^{ i( \Phi_{\rm i j } - \Phi_{\rm L'} ) } +  {\Omega_{  \rm i j}^{\rm L'}}^\ast e^{ i( \Phi_{\rm i j } + \Phi_{\rm L'} ) } \right]
\end{eqnarray*}
	The Rabi frequency
  $\Omega_{  \rm i j}^{\rm L} $ characterizes the strength of the transition between the states $|\rm i\rangle$ and $|\rm j\rangle$ and is defined by 
	$$\hbar \Omega_{\rm i j}^{\rm L} = \langle {\rm i} |  { \hat {\bm  \mu}}.{\bm E}_{\rm L} | {\rm j }\rangle  e^{i  {\bm k}_{\rm L}. {\bm r}}$$
		For more generality we have
written these equations for
$ \tilde \rho_{\rm i j} = \rho_{\rm i j}  {\rm e}^{ i \Phi_{\rm i j} (t) }$, where $\Phi_{ \rm i j} (t)$  would have to be later chosen cleverly for instance to remove the time dependent exponential terms ($\Phi_{ \rm i i} = 0$).

The next step \cite{2004PhRvA..69f3806B} consists to formally integrate the second equation 
	$\dot{	  \tilde \rho}_{\rm i j}  - b(t) {\tilde \rho}_{\rm i j} = \sum_{\rm L'} c_{\rm L'}(t)$ in
${\tilde \rho}_{\rm i j} (t)= \sum_{\rm L'} \int_0^t c_{\rm L'}(t') e^{\int_{t'}^t b} {\rm d} t'$. The standard choice of $\Phi_{\rm i j} = \Phi_{\rm L}$ is not possible when several lasers $\rm L$  are present, thus we report the formula in the first equation which then does not depend on the choice of $\Phi_{\rm i j}$.
We finally 
 make a statistical average (over several experiments) noted
$\uprho_{\rm i j} = < \rho_{\rm i j } > $.

The situation may be complex with, for instance, several lasers coherently driving the same transition to realize a travelling optical lattice plus incoherent lasers on the same or other transitions for optical pumping.
	Following	\cite{2004PhRvA..69f3806B},
we shall model the phase $\Phi^{\rm L} (t)$ as a Poissonian  random
process with $T_{\rm L} = 2/\Gamma_{\rm L}$ as the average time between successive phase jumps. 
The statistical average $< e^{ i[ \Phi^{\rm L} (t) -\Phi^{\rm L'} (t')] } >$ is non zero only if there is no phase jump between $t$ and $t'$. For the laser $\rm L$ this
occurs with a probability
$P_0 = e^{-(t-t')/T_{\rm L}}$. Thus, if $L$ and $L'$ are coherent, that is if
 $ \Phi_{\rm L} - \Phi_{\rm L'}
$ have no more fluctuating phases ($\Phi^{\rm L} = \Phi^{\rm L'}$), we have
$ < e^{i[ \Phi^{\rm L} (t) -\Phi^{\rm L'} (t')] } > =  e^{-(t-t') \Gamma_{\rm L}/2}$. 
The net result of this statistical average is to remove the $\Phi^{\rm L}$ fluctuating phases and simply to
add the laser linewidth $\Gamma_{\rm L}$, as done for the collisional terms  $\Gamma_{\rm coll}$, to the coherence relaxation rate which becomes half of
$$\Gamma^{\rm L}_{\rm i j} = \Gamma_{\rm coll}+\sum_{\rm k} (\Gamma_{\rm i k} + \Gamma_{\rm j k} )
	+\Gamma_{\rm L} $$

 \subsubsection{Rate equation}

Even if coherences, or Bloch equations, are tractable with a Monte Carlo method as described in \cite{2003PhRvA..67b2902M}, it is much simpler to use
 rate equations for each particle.  
	We thus use the method of steepest descent or stationary phase,  by integrating by part to write
	$\int_0^t {\rm d} t' \Omega_{\rm i j}^{\rm L}(t') e^{-[i(\omega_{\rm i j} \pm \omega_L) + \Gamma_{\rm i j}^{\rm L}/2](t-t') } \sim \frac{\Omega_{\rm i j}^{\rm L}(t)}{i(\omega_{\rm i j} \pm \omega_L) + \Gamma_{\rm i j}^{\rm L}/2 }  $. This is valid only for low saturation that is when the saturation parameter
		$$ s = \frac{2 |\Omega_{\rm i j}^{\rm L}|^2}{4(\omega_{\rm i j} - \omega_L)^2 + {\Gamma_{\rm i j}^{\rm L}}^2 } \ll 1$$
		The resonant terms, that is when
	$ | \delta_{\rm i j}^{\rm L,0} = \omega_{\rm L} - \omega_{\rm i j}  |\ll \omega_{\rm i j}$, are the dominant one and  
 terms containing $e^{i[\Phi_{\rm L} +\Phi_{\rm L'}]} $, will be dropped
(so called rotating-wave
approximation)   because they are much smaller than resonant ones and because 
the statistical average leads to exponential decreasing terms.
		A slow time variation 
due to small frequency difference between
coherent lasers $\rm L, L'$ driving the $\rm i j$ transition could still occur but, in order to avoid remaining oscillating terms, we shall suppose it smaller than $1/\Gamma^{\rm L}_{\rm i j}$.
We define the detuning
$$\delta_{\rm i j}^{\rm L} = \omega_{\rm L} - \omega_{\rm i j} - {\bm k}_{\rm L}.\bm v = \delta_{\rm i j}^{\rm L,0} - {\bf k_{\rm L}}.{\bf v} + \Delta_{\rm j} -  \Delta_{\rm i} $$
  where $\hbar \Delta_{\rm i}$ and $\hbar \Delta_{\rm j}$ are the energy shifts created by the external trapping potentials.
We combined  lasers in  class ``${\rm L_c}$" of mutually coherent lasers by writing $\sum_{\rm L} = \sum_{\rm L_c} \sum_{\rm L}^{\rm L_c} $.
An important parameter is the total Rabi Frequency for the class $\rm L_c$:
$$\Omega_{\rm i j}^{\rm L_c} = \sum_{\rm L}^{\rm L_c} \Omega_{\rm i j}^{\rm L}  e^{-i \delta_{\rm i j}^{\rm L} t} =  
\sum_{\rm L}^{\rm L_c} \langle {\rm i} |  { \hat {\bm  \mu}}.{\bm E}_{\rm L} | {\rm j }\rangle  e^{i  {\bm k}_{\rm L}. {\bm r} (t)} e^{-i( \omega_{\rm L} - \omega_{\rm i j})t}$$
Where  $ {\bm r} (t) = \bm r + \bm v t$, which in the simulation is taken to be the current location of the particle.

 We finally obtain the rate equations  used in our simulation and illustrated in Fig. \ref{fig:two_level}:
\begin{eqnarray}
  	\dot	 \uprho_{\rm i i} & = &  \sum_{\rm j} \left[ \Gamma_{\rm j i}  \uprho_{\rm j j}  
			-  \Gamma_{\rm i j}  \uprho_{\rm i i}  + \gamma_{\rm i j} (\uprho_{\rm j j} -\uprho_{\rm i i} )
		\right] \label{eq_rate}\\
	\gamma_{\rm i j} &=&\sum_{\rm L_c} \frac{| \Omega_{\rm i j}^{\rm L_c} |^2}{{\Gamma^{\rm L_c}_{\rm i j}}^2+4 {\delta_{\rm i j}^{\rm L_c}}^2} \Gamma^{\rm L_c}_{\rm i j}	 \label{rate_exc}
\end{eqnarray}
The physical interpretation is obvious,
$\gamma_{\rm i j}$ is the rate of excitation, but also of stimulated emission, of the transition and is given simply by the incoherent sum over the individual rates. 

With $\langle {\rm i} |  { \hat {\bm  \mu}}.{\bm E}_{\rm L} | {\rm j }\rangle = 
|\langle {\rm i} |  { \hat {\bm  \mu}}.{\bm e}_{\rm p} | {\rm j }\rangle  E_{\rm L} \epsilon_p| e^{ i  \phi^{\rm L}_{\rm p}} $  we see that the rate is proportional to the total laser taken into account interferences due to lasers polarizations:

\begin{align}
| \Omega_{\rm i j}^{\rm L_c} |^2 & =  \sum_{\rm L}^{\rm L_c} |\Omega_{\rm i j}^{\rm L}|^2 +  \label{Rabi} \\
\MoveEqLeft 2 \sum_{\rm L' > L}^{\rm L_c} |\Omega_{\rm i j}^{\rm L} \Omega_{\rm i j}^{\rm L'}|
\cos[(\omega_{\rm L} - \omega_{\rm L'})t - ( {\bm k}_{\rm L} - {\bm k}_{\rm L'}). {\bm r} (t) - (\phi^{\rm L}_{\rm p} - \phi^{\rm L'}_{\rm p}) ]  \nonumber
\end{align}
where $|\hbar \Omega_{\rm i j}^{\rm L} | = |\mu \epsilon_p E_{\rm L }|   $.

 \subsubsection{Bloch equation with laser linewidth}

For completeness, we mention that
in the case of a single class of coherent lasers ${\rm L_c}$, we can choose one laser $\rm L_0$ of this class and
$\Phi_{\rm i j} = \Phi_{\rm L_0}$ to derive generalization of the Bloch equation. Time derivative of the statistical average equations, before the
integration by part,
leads to  Bloch equations with laser linewidth:
	\begin{eqnarray*}
	  	\dot	 \uprho_{\rm i i}  & = &  \sum_{\rm j} \Gamma_{\rm j i}  \uprho_{\rm j j}  
			- \sum_{\rm j} \Gamma_{\rm i j}  \uprho_{\rm i i}  
			- \sum_{\rm j} {\cal I}m \left(\tilde \uprho_{\rm j i} \Omega_{\rm i j} \right)  \\
	\dot{	  \tilde \uprho}_{\rm i j}  & = &  i \delta_{\rm i j}^{\rm L_c} {\tilde \uprho}_{\rm i j} -\frac{
	\Gamma^{\rm L_c}_{\rm i j}  }{2}   		
{\tilde \uprho}_{\rm i j}  + 
	\frac{i}{2}  ( \uprho_{\rm j j} -  \uprho_{\rm i i}  ) \Omega_{\rm i j}
\end{eqnarray*}
where $\Omega_{\rm i j }= \sum_{\rm L}
 {\Omega_{  \rm i j}^{\rm L}} e^{ i( \Phi_{\rm i j } - \Phi_{\rm L} ) }$.
We obviously recover the rate equation (\ref{eq_rate}) using $	\dot{	  \tilde \uprho}_{\rm i j}  =0$ and we find
$
 { \tilde \uprho}_{\rm i j} = \frac{ \Omega_{\rm i j}}{2 \delta_{\rm i j}^{\rm L_c} + i \Gamma^{\rm L_c}_{\rm i j}} (\uprho_{\rm  i i} - \uprho_{\rm j j})
$
which, as expected, has a smaller value, by the saturation parameter $s$, than the populations $\uprho_{\rm  i i} - \uprho_{\rm j j}$.

 \subsubsection{Dipolar potential and light shift}

A full quantum theory of light shift and heating in an optical trap can be found in Ref \cite{2010PhRvA..82a3615G}, but we shall here 
follow a simple deviation based on Ehrenfest's theorem. Atomic motion is affected by the force
$\bm F  = \langle  \hat{\bm \mu} \rangle  .  \bm \nabla  {\bm E} $ where
$ \langle  \hat{\bm \mu} \rangle = Tr( \hat {\bm \rho} \hat{ \bm \mu} )  = \sum_{\rm i j}  \hat {\bm \rho}_{\rm j i} \hat{ \bm \mu}_{\rm i j}$. 
 We first write it in the single coherent lasers class case, because general result is simply the sum over the classes.
 The statistical average leads to
${\rm {\bm  F}}  =  \hbar \sum_{\rm i > j} {\cal R}e \left\{  {\tilde \uprho}_{\rm i j}^\ast  \bm \nabla \Omega_{  \rm i j} 	  \right\}   $ where we have separated  the sum using levels with $\omega_{\rm i} > \omega_{\rm j}$. This can be written as:
${\rm {\bm  F}}  =  -\sum_{\rm i > j}(\uprho_{\rm  j j} -\uprho_{\rm i i}) \hbar \gamma_{\rm i j} \left[ \frac{2 \delta_{\rm i j}}{\Gamma_{\rm i j}^{\rm L_c} }  {\cal R}e \left\{ \frac{ \bm \nabla \Omega_{\rm i j} }{\Omega_{\rm i j} } \right\}
-  {\cal I}m \left\{ \frac{ \bm \nabla \Omega_{\rm i j} }{\Omega_{\rm i j} } \right\}
\right] $ where a clear separation between the dipolar and the scattering force appear ${\rm {\bm  F}} ={\rm {\bm  F}_{\rm dip}} +{\rm {\bm  F}_{\rm scat}} $.

	In the general case we found:
	$${\rm {\bm  F}_{\rm dip}}  	= - \bm \nabla U_{\rm dip}  \ \ ; \ \ U_{\rm dip} =  \sum_{\rm L_c} \sum_{\rm i>j}  \hbar  \gamma_{\rm i j}(\bm r) \frac{\delta_{\rm i j}   }{\Gamma_{\rm i j}^{\rm L_c}} (\uprho_{\rm j j}  -   \uprho_{\rm i i} ) $$	
		In our Monte Carlo simulation, a molecule is always in a given state  $| j \rangle$. 
		Thus, in addition to the previous shifts created by the magnetic and electric fields, the dipolar potential indicates that we have to shift the $j$ energy level by
		\begin{equation} 
	 \sum_{\rm L_c} \sum_{\rm i>j}	\hbar \delta_{\rm i j}^{\rm L_c}  \frac{ \gamma_{\rm i j}^{\rm L_c}(\bm r) }{\Gamma_{\rm i j}^{\rm L_c}} - 	 \sum_{\rm L_c} \sum_{\rm k< j}	\hbar \delta_{\rm k j }^{\rm L_c}  \frac{ \gamma_{\rm k j}^{\rm L_c}(\bm r) }{\Gamma_{\rm k j}^{\rm L_c}}
		\end{equation}
	We recover the standard result for a two level system that for red detuning the upper state is up-shifted and the lower one is down-shifted by the laser.
	
		This picture is not the same that the usual dressed state one \cite{1985JOSAB...2.1707D} where particles are in dressed state and in a superposition of
		 $| i \rangle$ and $| j\rangle$ levels with a light shift given by $\hbar \delta \ln (1 + s)$. However we recover the same average shift  in our Monte Carlo simulation because we oscillate in time between  $| j \rangle$ and $| i\rangle$. 
Finally, we mention that we use this shift only to calculate the forces but we do not add it as a real shift in the levels. The energy levels are thus not shifted by lasers and no 
		   new laser detuning are calculated. The main reasons is  to avoid unphysical accumulation of dipole shifts. Indeed, if we calculate a new detuning with this formula, at the next update of the energy levels a novel detuning will be added to this one and we would have accumulation of the detuning that would not correspond any-more to reality.  
			
		The scattering force is given by
		\begin{align*}
	{\rm {\bm  F}_{\rm scat}} & =  \sum_{\rm L_c} \sum_{\rm i>j}  \hbar  \gamma_{\rm i j}^{\rm L_c}  (\uprho_{\rm j j}  -   \uprho_{\rm i i} )  {\cal I}m \left\{ \frac{ \bm \nabla \Omega_{\rm i j}^{\rm L_c} }{\Omega_{\rm i j}^{\rm L_c} } \right\}   \\
 | \Omega_{\rm i j}^{\rm L_c} |^2 {\cal I}m \left\{ \frac{ \bm \nabla \Omega_{\rm i j}^{\rm L_c} }{\Omega_{\rm i j}^{\rm L_c} } \right\}  
& = 
 \sum_{\rm L}^{\rm L_c} \bm k_{\rm L} |\Omega_{\rm i j}^{\rm L}|^2 +  2 \sum_{\rm L' > L}^{\rm L_c} \frac{\bm k_{\rm L} + \bm k_{\rm L'}}{2} \times  \\
\MoveEqLeft[9]|\Omega_{\rm i j}^{\rm L} \Omega_{\rm i j}^{\rm L'}|
\cos[(\omega_{\rm L} - \omega_{\rm L'})t - ( {\bm k}_{\rm L} - {\bm k}_{\rm L'}). {\bm r} (t) - (\phi^{\rm L}_{\rm p} - \phi^{\rm L'}_{\rm p})]
\end{align*}
 The photon absorption rate is  given	by $ \gamma_{\rm i j}^{\rm L_c} $ and the photon momentum can be chosen depending on the ``participation" of the laser beams to the total irradiance by comparison with formula \ref{Rabi}.
	In the simple example of a retro-reflected laser creating a standing wave lattice ($\Omega_{\rm i j}^{\rm L} = \Omega_{\rm i j}^{\rm L'}$ and $\bm k_{\rm L} = - \bm k_{\rm L'}$) no force is present but the absorption rate is non zero.
	
More general cases,  especially in the saturated regime, can be treated with ad-hoc formulas (see \cite{Woh1}) but we have not implemented them.

\subsubsection{General laser spectrum}

We have, up to now, followed a standard procedure to derive the absorption rate (\ref{eq_rate}): that is statistical average (and deconvolution approximation) followed by an integration by part 
\cite{2004PhRvA..69f3806B,2012AmJPh..80..882H}. However, the result seems non-physical in the far-off resonance case, when $|\delta_{\rm i j}^{\rm L_c}|\gg \Gamma^{\rm L_c}_{\rm i j}$. We found an absorption rate that is proportional to the laser linewidth, where obviously, because the laser is very far from resonance, only the laser intensity, not its linewidth, should play a role \cite{Gri1}. 
The problem arises in a Cauchy-lorentzian laser irradiance spectrum because the spectrum does not decrease fast enough and has still a large amplitude even far from its center. But, the problem is more general and even for other laser spectrum
it arises  from the fastest oscillating term in $e^{-[i ( \delta_{\rm i j}^{\rm L} + \Phi^{\rm L}) + \Gamma_{\rm i j}^{\rm L}/2](t) }$ that is now due to the detuning and not any more to the laser phase fluctuation. Thus, in such far detuned case the  integration by part has to be performed before the statistical average. The net result, for this far off resonant case, is that the phase fluctuation plays no more role and thus $\Gamma_{\rm L}$ has to be removed from
$\Gamma^{\rm L}_{\rm i j}$ which becomes $ \Gamma^{\rm L}_{\rm i j} = \Gamma_{\rm coll}+\sum_{\rm k} (\Gamma_{\rm i k} + \Gamma_{\rm j k} )
= \Gamma^{\rm i j} $ in the absorption rate formula (\ref{rate_exc}).

Fortunately, the near and far off resonant results can be combined together in a single formula.
Indeed  Eq.  (\ref{rate_exc}) can be written,
 in the single laser case to simplify, as $ \gamma_{\rm i j}  =  L_{\upomega} \otimes I_{\upomega} (\omega_{\rm i j} + {\bm k}_{\rm L}.\bm v) \uppi |\mu \epsilon_p|^2  / (\epsilon_0 c)$ that is proportional to the convolution of the  Cauchy-Lorentzian natural linewidth $L_{\upomega}(\omega) =  \frac{1}{\uppi} \frac{\Gamma^{\rm i j}/2}{\omega^2 + (\Gamma^{\rm i j}/2)^2} $ by a
 Cauchy-Lorentzian laser irradiance spectrum $I_{\upomega} (\omega)$. 
 However
 for more realistic laser spectral shape such as a Gaussian one  the formula 
\begin{equation}
\gamma_{\rm i j}  =  [L_{\upomega} \otimes I_{\upomega}] (\omega_{\rm i j} + {\bm k}_{\rm L}.\bm v ) \uppi |\mu \epsilon_p|^2  / (\epsilon_0 c) \label{eq_rate_exc}
\end{equation}
 agrees with our expectation  even in the far-off resonant case where the formula agrees with the well known result $\gamma_{\rm i j}   =   \frac{|\Omega_{\rm i j}^{\rm L}|^2}{4 {\delta_{\rm i j}^{\rm L}}^2} \Gamma^{\rm i j} $.
As another example, for very broadband lasers with a spectrum  including the resonance, the Lorentzian natural linewidth can be considered as a Dirac peak and $\gamma_{\rm i j} $ is simply proportional to the irradiance at the transition wavelength $I_{\upomega} (\omega_{\rm i j} + {\bm k}_{\rm L}.\bm v)$. 
 The physical interpretation of Eq. (\ref{eq_rate_exc}) is obvious: in an incoherent (rate equation) model, a laser lineshape can be seen as a sum of several narrowband (delta or narrow Gaussian) lasers lines of intensity $I_{\upomega} (\omega)$,  so, in a low saturation and perturbative approach, the total rate of excitation is thus simply the sum of the incoherent rates.

In our simulation, we  keep the same spirit, and for each coherent lasers class, without making a precise integration of the formulas over the laser spectrum,
we calculate the rates using Eq. (\ref{eq_rate_exc}) and  then forces  simply  by replacing $ \Gamma^{\rm L}_{\rm i j} $ by
$ \Gamma^{\rm i j} $ in the formulas.

 \subsubsection{Parameters needed}

\label{param_need}

We want here to discuss the irradiance needed to efficiently perform the transition. This is usually defined by the saturation irradiance (intensity) $I_{\rm sat}$ defined as $\gamma = \frac{I}{I_{\rm sat}} \frac{\Gamma}{2}$ (where we have dropped the ij subscript).
 With a laser linewidth at least as large as the transition linewidth we have $\Gamma_{\rm i j}^{\rm L} \approx \Gamma_{\rm L}$ and $\gamma \approx \frac{\Omega^2 \uppi}{2} \frac{I_{\upomega} (\omega_{\rm i j})}{I}$. At resonance 
$\gamma \sim \frac{\mu^2 E_L^2 \Gamma_{\rm L}}{ \hbar^2 \Gamma_{\rm L}^2} $. Including the Franck-Condon (FC) and angular factors (AF), Einstein's formula for the natural lifetime $1/\Gamma$ leads to
$\gamma/{\rm FC.AF} \sim  \frac{6 \uppi  c^2 }{\omega^3 \hbar} I \frac{\Gamma}{\Gamma_{\rm L}}$ and so
$$ \gamma/{\rm FC.AF} \sim 300 \, 000 \, s^{-1} \left( \frac{\lambda}{500 \, {\rm nm}}  \right)^3 \frac{I}{{\rm W/m}^2} \frac{\Gamma}{\Gamma_{\rm L}} $$

Depending on the laser transition, the needed transition rate is not the same. Clearly for the Sisyphus transition the transition probability  should be near unity when the particle arrive at the transition point; whereas for the repumping transition we may have much more time to do it (except if it is needed to repump the particle during the Sisyphus transfer), typically a fraction of the trapping period.

For the narrow linewidth Sisyphus laser, in order to have an efficient transition,
	the time $\gamma^{-1}$ to realize the transition should match the time where the particle spend at resonance (that is when $|\delta| < \Gamma_{\rm L}$).
 Due to the (thermal) velocity $v = \sqrt{{\rm k}_{\rm B} T / m} \sim 1-10$ m/s in the trap (depth ${\rm k_B} T$ and size $\sigma_r \sim 1$ mm),  the time spend at the resonance condition is only $\frac{ \hbar \Gamma_L}{{\rm k_B} T} \frac{\sigma_r}{v}$. 
We thus find 
$ \gamma \sim  \frac{k{\rm _B} T}{ \hbar \Gamma}
\frac{v}{\sigma_r}$. 
 For an harmonic trap $\sigma_r \propto  \sqrt{T/\omega_1}$ and the irradiance needed  is thus proportional to the temperature. This explain why we have to reduce the power during the cooling process.
As example, a (150 ns lifetime, 500nm wavelength) 1 MHz  transition natural linewidth require 
$\gamma \sim  10^7 $s$^{-1}$ and an irradiance of
$I \sim  100 $W/m$^2$.

For the case of a laser used for optical (re-)pumping, we can simply choose its broadband linewidth $\Gamma_L$ such that the particle is always at resonance, i.e. $\hbar \Gamma_L\sim {\rm k_B} T$. In practice we use a much broader, and thus a correspondingly more intense, laser in order to cover several rotational or vibrational levels.
For $T\sim 100$ mK this require a linewidth of $\Gamma_L \sim 2\uppi $(2 GHz) and for a transition rate of $\gamma^{-1} \sim 1$ ms an irradiance of 7 mW/cm$^2$.

\subsection{Kinetic Monte Carlo exact solution of the master or rate equation}
\label{KMC_model}

	We recall  here the method, which have been published in Ref. \cite{2008NJPh...10d5031C}, used to perform the simulation and especially the Kinetic Monte Carlo method for solving rate equations and the N-body integrator.
	
	The method is able to solve any system  driven by a master equation
\begin{equation}
    \frac{dP_k}{dt} =  \sum_{l=1}^N \Gamma_{k l} P_l - \sum_{l=1}^N  \Gamma_{l k}  P_k 
    \label{mastereq}
\end{equation}
This equation
describes the time evolution of the probability $P_k$ of a system to occupy each one of a discrete set of states numbered by $k$.  Each process occurs at a certain average rate $\Gamma_{l k} (t)$.

With, $P_{\rm k} = \uprho_{\rm k k}$ we clearly see that our system is described with such a master equation. The states ensemble contains all molecule internal states and the rate are the  the spontaneous emission plus the lasers absorption and stimulated emission  rates described in Eq (\ref{eq_rate}).

One of the simplest algorithm to solve this equation,  sometimes called fixed time step algorithm,  is based on the first-order formula 
$ P_k(t+dt) \approx P_k(t) + \sum_{l=1}^N \Gamma_{k l}(t)  P_l (t) dt - \sum_{l=1}^N  \Gamma_{l k}(t) P_k(t) dt    $.
The main disadvantage of this fixed time step method \cite{2003cond.mat..3028J} is that  $d t$ has to be small enough to maintain accuracy and such that at most one reaction occurs during each time step: meaning  $\Gamma_{l k}(t)  dt \ll 1$. 
 The Kinetic Monte Carlo  (KMC) algorithm 
 solve this problem by  choosing optimal time step  evolution of the system. Furthermore, the KMC method makes exact numerical calculations and cannot be distinguished from an exact molecular dynamics simulation, but is orders of magnitude faster. KMC method  is indistinguishable from
the behavior of the real system, reproducing for instance all possible data in an experiment including its statistical noise. 
Surprisingly enough, up to now it has been
more or less limited to the study of
chemical reactions, surface or cluster physics (diffusion, mobility, vacancy motion, transport process, epitaxial growth, dislocation, coarsening,  ...).

The Kinetic Monte Carlo  algorithm uses the fact that the system has a
 Markovian behavior.
For a
system initially at time $t$ in state $k$,
the probability that the system
has not yet escaped from state $k$ at time $t'$ is given by
$ \exp \left( \int_t^{t'} \sum_l \Gamma_{l k}(\tau) d \tau  \right) $.
At time $t'$ a reaction takes place, so just before $t'$ the system is still in state $k$,
we therefore have to
generate a new configuration $l$ by picking it out of all possible new configurations with a probability proportional to
$\Gamma_{lk} (t')$. 

The KMC algorithm is then the iteration of the following steps.
\begin{itemize}
\item Initializing the system to its given state called $k$ at the actual time $t$.
	\item Creating the new rate list $ \Gamma_{l k} $ for the system, $l=1,\ldots,N$.
	\item  Choosing a unit-interval uniform random number generator \footnote{In our case we use the free 
 implementations by GSL (GNU Scientific Library) of the Mersenne twister unit-interval uniform random number generator  of Matsumoto and
Nishimura.} $r$: $0<r\leq 1$ and  calculating the first reaction rate time $t'$ by solving
	$ \int_{t}^{t'} \sum_{l=1}^N \Gamma_{l k} (\tau) d \tau =  - \ln r$.
	\item Choosing a unit-interval uniform random number generator  $r'$: $0<r'\leq 1$ and searching for the integer $l$ for which $R_{l - 1} < r' R_N \leq  R_l$ where
 $R_j = \sum_{i=1,j} \Gamma_{i k} (t')$ and $R_0=0$. This can be done efficiently using a binary search algorithm.
	\item Setting the system to state $l$ and modifying the time to $t'$. Then go back to the initial step.
	
\end{itemize}

\subsection{Simple method to solve the N-body equations of motion}

Discussion of possible solutions to solve the N-body equations of motion is given in Ref. \cite{2008NJPh...10d5031C}. We simply recall here the main conclusions.

Numerical methods such as the ordinary Runge--Kutta methods are not ideal for integrating Hamiltonian systems because they do not conserve energy. On the contrary symplectic integrators  does conserve energy. 

Because of the computational round-off error and  stability domain issues, algorithms are generally not good to go beyond third or fourth order. One very popular N-body integrator is the fourth (local) order ``Hermite'' predictor-corrector scheme by Makino and Aarseth \cite{heggie2003gravitational,Aarseth,2008NewA...13..285P}. We use here
a simpler but still efficient algorithm, the so called velocity leapfrog-Verlet-St\"ormer-Delambre algorithm:

\begin{eqnarray}
\mathbf{r} (t+ \Delta t) &=& \mathbf{r} (t) + \mathbf{v} (t) \Delta t + \frac{1}{2}
\mathbf{a}(t) (\Delta t)^2 \label{eq:motion} \\
\mathbf{v} (t+ \Delta t) &=& \mathbf{v} (t) + \frac{1}{2} \left(
\mathbf{a}(t) + \mathbf{a}(t+\Delta t)  \right)(\Delta t) \nonumber
\end{eqnarray}
It is of  $O((\Delta t)^3)$ accuracy for both position $\mathbf{r}$ and velocity $\mathbf{v}$ for a $\Delta t$ timestep.
It has also the big advantage that accuracy can be improved by using  higher order symplectic integrators such as the one by Yoshida
\cite{1990PhLA..150..262Y}.

In our case the velocity can be modified directly by photon recoils in absorption or in emission and the acceleration is simply calculated using the gradient of the  potential.
 A typical time-scale is given by the motion in laser (or trapping) fields. Thus the time-step should be small fraction of the ratio of the waist, or the trap size, over the thermal velocity of the particle.

Finally, we  combine the KMC and the N-body integrator in the following way: we first calculate an expected (KMC) reaction time $t'-t$. Then, if $t'-t < \Delta t$, that is a reaction should occurs before the motion,  we move, through Eq. (\ref{eq:motion}) the particles, using the timestep $t'-t$ and the reaction takes place at time $t'$. On the contrary, if $\Delta t < t'-t$,
a dynamical (N-body) evolution of the system is made, no reaction takes place but due to its markovian probabilistic behaviour, the system is still govern by Eq. (\ref{mastereq}). Then,
after each change, of the position or of the internal state, the laser fields and potentials are recalculated and new transition rates are calculated. 
It is convenient  to choose a  $\Delta t$ timestep such that the calculated laser excitation  rates are almost constant over it, in order
to allow us to calculate the reaction time $t'-t = - \frac{\ln r}{\sum_{l=1}^N \Gamma_{l k} (t)}$.
We use these rules of thumb to start the simulation but we finally 
reduce  the timestep $\Delta t$ until we obtain convergence of the results which usually occurs for $\gamma \Delta t \sim 1$.

\section{List of formed cold molecules}
\label{list:coldMol}

  The following table presents  a list of the cold molecular species with $T<1$~K. For each species only the reference corresponding to the pioneering work  is given. 
The  combination of the  buffer gas pre-cooling  with the Stark velocity filtering is listed either under the ``Cryogeny" or under the ``Velocity filtering" category.
  The occurrence of some ``exotic" molecules has been detected only as a loss process in a degenerate gas but not really produced and isolated. This is the case of Cs$_3$ and
 Cs$_4$~\cite{2005PhRvL..94l3201C} and thus they do not appear in the table.  Similarly, several photoassociation experiments have produced electronically excited cold molecules (as single example Yb~\cite{2004PhRvL..93l3202T}) but we do not include them in the list as they have a lifetime limited to some tens of nanoseconds. Furthermore, we do not always detail the isotope difference, for instance
  NH$_3$ can be the same as $^{15}$NH$_2$D.

\cleardoublepage 
\begin{table}[htbp]
{ 
\footnotesize
\begin{tabular}{|l|l|c|c |}
  \hline
   Method &  Molecule  & T($\mu$K)  & N \\
  \hline
    \hline
 Feshbach, RF \cite{2010RvMP...82.1225C,2006RvMP...78.1311K} & 
 \begin{tabular}{l} \\
 $^{85,87}$Rb$_2$ \cite{2002Natur.417..529D,2005PhRvL..95s0404T,2006PhRvL..97r0404P}, 
 Cs$_2$ \cite{2003Sci...301.1510H},\\
  $^{40}$K$_2$ \cite{regal2003}, 
    Li$_2$ \cite{2003Sci...302.2101J,2003PhRvL..91x0401C,2003PhRvL..91h0406S},\\ 
 Na$_2$ \cite{xu2003formation}, 
 $^{40}$KRb \cite{2006PhRvL..97l0402O},\\
   $^{41}$K$^{87}$Rb   \cite{2008PhRvA..78f1601W},
 Cr$_2$ \cite{2009PhRvA..79c2706B},
 Li$_3$ \cite{2010Sci...330..940L}, NaK \cite{2012PhRvL.109h5301W}
 \end{tabular}
 & 0.1 & 100 000 \\
  \hline
  Photoassociation \cite{2006RvMP...78..483J} & 
  \begin{tabular}{l}\\
  Cs$_2$ \cite{1998PhRvL..80.4402F}, 
    H$_2$ \cite{1999PhRvL..82..307M}, 
  Rb$_2$ \cite{2000PhRvL..84.2814G}, \\
  Li$_2$ \cite{2000Natur.408..692G}, 
  Na$_2$ \cite{2002PhRvA..66e3401F}, \\
  K$_2$ \cite{nikolov2000,2003Natur.426..537G,2007PhRvL..98t0403G}, 
  He$^*_2$ \cite{2006PhRvL..96b3203M},\\
  Ca$_2$ \cite{2000PhRvL..85.2292Z},
   KRb \cite{2004PhRvL..93x3005W},
   RbCs \cite{2004PhRvL..92o3001K},\\
   NaCs \cite{2004PhRvA..70b1402H},
   LiCs \cite{2008PhRvL.101m3004D}
   \end{tabular}
   & 100 & 200 000 \\
  \hline
  Three body collision &\begin{tabular}{r}\\
  Rb$_2$ \cite{2000PhRvL..84.2814G}
  Li$_2$\cite{2003Sci...302.2101J,2003PhRvL..91y0401Z}
  \end{tabular}
  & 0.2 & 2 000 000 \\
   \hline
	\hline
	  laser cooling & SrF \cite{shuman2010laser}, YO \cite{2013PhRvL.110n3001H}, CaF \cite{2013arXiv1308.0421Z}
  & 0.3 &  \\
	\hline
       \hline
    Cryogeny (Buffer Gas) \cite{2011arXiv1111.2841H} &
     \begin{tabular}{l} \\
    CaH \cite{1998Natur.395..148W},
    VO \cite{1998JChPh.109.2656W}
    CaF \cite{2005PhRvL..94l3002M},\\
    PbO \cite{2001PhRvA..63c0501E,2005PhRvL..95q3201M}, 
    O$_2$  \cite{2007JChPh.126o4307P} 
    NH \cite{2004EPJD...31..307E},\\
    ND,
    CrH, MnH \cite{2008PhRvA..78c2707S},
    ND$_3$, \\
     H$_2$CO \cite{2009PhRvL.102c3001V}, 
    YbF \cite{2009NJPh...11l3026S}, \\
    NH, NH$_3$, O$_2$, \\
    ThO, Naphthalene \cite{2010APS..MARH24001D}, \\
		BaF, SrO, YbF, YO \cite{2011arXiv1111.2841H}
    \end{tabular}
    & 400 000 & 10$^{12}$ \\
        \hline
         Field Slowing: Stark \cite{2008NatPh...4..595V} & 
    		\begin{tabular}{l}\\
			CO \cite{1999PhRvL..83.1558B},
    NH$_3$,
    ND$_3$ \cite{2000Natur.406..491B,2001Natur.411..174C},\\
    OH \cite{2003PhRvL..91x3001B,2004PhRvA..70d3410B}, 
    OD \cite{2005PhRvL..94b3004V},
    H$_2$CO \cite{2006PhRvA..73f3404H},\\
    NH \cite{2006JPhB...39.1077V}, 
    SO$_2$ \cite{2008EPJD...46..463B}, 
    C$_7$H$_5$N  \cite{2008PhRvA..77c1404W},\\
    YbF  \cite{2004PhRvL..92q3002T}, \\
    LiH \cite{2009NJPh...11e5038T}, 
    CaF \cite{2010PhRvA..81c3414W}
		\end{tabular}
    & 10 000 & 1 000 000  \\ 
    Rydberg &
    H$_2$ \cite{2004PhRvL..92c3005V,2007JPhCS..80a2045Y,2008PhRvL.100d3001H,2009PhRvL.103l3001H}
    &  &  \\
    Optical &
    C$_6$H$_6$  \cite{2004PhRvL..93x3004F}, NO  \cite{2006JPhB...39.1097F} &  & \\
    Zeemann &
    O$_2$ \cite{2008PhRvA..77e1401N} 
    &  &  \\
    \hline
        Beam collision& \begin{tabular}{l}\\
    NO \cite{2003Sci...302.1940E}, 
    KBr (13K) \cite{2007PhRvL..98j3002L},\\
    ND$_3$ \cite{2009FaDi..142..143K} \end{tabular}
    & 400 000 & \\
    Beam dissociation &
   NO \cite{2010arXiv1002.3698T} &
   1 600 000 & \\
    \hline
    Rotating Nozzle & \begin{tabular}{l}\\
    O$_2$,
    CH$_3$F,\\
    SF$_6$ \cite{gupta1999,gupta2001},
    CHF$_3$ \cite{2010PhRvA..81c3409S},\\
      perfluorinated C$_{60}$  \cite{2008EPJD...46..307D}
      \end{tabular}
    & 1 000 000 & \\
        \hline
    Velocity filtering 
&
\begin{tabular}{l}\\
    H$_2$CO \cite{2006PhRvA..73f3404H},
    ND$_3$ \cite{2003PhRvA..67d3406R}, \\
    D$_2$O  \cite{2006PhRvA..73f1402R},
    CH$_3$F \cite{2008PhRvL.100d3203W},  \\
	CF$_3$H \cite{2009FaDi..142..203S},
    CH$_3$CN \cite{2010PCCP...12..745L}, H$_2$O , D$_2$O,\\  
       HDO \cite{2009PhRvA..79a3405M}, NH$_3$,  CH$_3$I, \\
       C$_6$H$_5$CN,  C$_6$H$_5$Cl  \cite{2010JPhB...43i5202T}
    \end{tabular}
    & 1 000 000 & 10$^9$ \\
    \hline
    Sympathetic cooling \cite{2009JPhB...42o4001W,low_temp_Mol_2008,cold_Mol_2009} &
    	\begin{tabular}{l}\\
     BeH$^+$, YbH$^+$ \cite{1997quant.ph.10025W},  \\
     AF350$^+$=C$_{16}$H$_{14}$N$_2$O$_9$S$^+$ \cite{2006PhRvL..97x3005O}, \\
      MgH$^+$ \cite{2000PhRvA..62a1401M}, O$_2^+$, MgO$^+$, CaO$^+$ \cite{phdHornekaer2000},  \\
      H$_2^+$,  H$_3^+$ \cite{2005PhRvL..95r3002B}, BaO$^+$ \cite{2008PhRvA..78d2709R}, NeH$^+$, \\
      N$_2^+$, OH$^+$, H$_2$O$^+$, HO$_2^+$, \\
    ArH$^+$, CO$_2^+$, KrH$^+$, \\
    C$_4$F$_8^+$, R6G$^+$ \cite{cold_Mol_2009}, \\
        Cyt$^{12+}$,  Cyt$^{17+}$ \cite{2008PhRvA..78f1401O} \\
        GAH$^+$=C$_{30}$H$_{46}$O$_4$ \cite{2009JPhB...42c5101O} 
    	\end{tabular}
    & 20 000 & 1000 \\
    \hline
		Nanodroplet \cite{2000JChPh.112.8068E,choi2006infrared,2006JPhB...39R.127S,2006JLTP..142....1B,2007IRPC...26..249K,Handbook_spectro} &
    	\begin{tabular}{l}\\
		 Mg$_{1-3}$HCN, Ne$^{-}$, Kr$^{-}$, ArHF, tetracene-Ar, \\
		HCN-H$_2$, HD, D$_2$, Ag8-Ne$_N$, Ar$_N$, Kr$_N$, Xe$_N$ ($N = 1-135$), \\
		NaCs, LiCs, HF-(H$_2$), OCS-(H$_2$)$_N$ ($N = 1-17$), \\
		Cs$_2$, Rb$_2$, CO, HCl, amino acid, 3-hydroxyflavone, \\
	xanthine, [Na(H$_2$O)$_N$]$^+$ ($N = 6-43$) \cite{2010RScI...81e4101F}, HF-N$_2$O, Mg-HF, \\
	 Mg-(HF)$_2$, CH$_3$-H$_2$O, Cs$_N^+$(H$_2$O)$_M$,	CF$_3$I, CH$_3$I, ...
    \end{tabular} & 1 000 000 & 10 \\
    \hline
		\end{tabular}
}
\vspace{.5cm}
\caption{Slow and cold molecule list with $T<1$~K with an estimation of their temperature $T$ and number $N$. For each species only the reference corresponding to the pioneering work  is given. }
\label{lis_coldMolec}
\end{table}
\cleardoublepage

\bibliographystyle{h-physrev}

\begin{thebibliography}{}

\end{thebibliography}


\begin{thebibliography}{100}

\bibitem{2000PhRvA..62a1401M}
K.~{M{\o}lhave} and M.~{Drewsen},
\newblock \pra {\bf 62}, 011401 (2000).

\bibitem{1981PhRvL..46..236D}
N.~{Djeu} and W.~T. {Whitney},
\newblock Physical Review Letters {\bf 46}, 236 (1981).

\bibitem{2009FaDi..142..191S}
P.~{Sold{\'a}n}, P.~S. {Uchowski}, and J.~M. {Hutson},
\newblock Cold and Ultracold Molecules, Faraday Discussions, volume 142, 2009,
  p.191 {\bf 142}, 191 (2009), 0901.2493.

\bibitem{2012Natur.492..396S}
B.~K. {Stuhl} {\em et~al.},
\newblock \nat {\bf 492}, 396 (2012), 1209.6343.

\bibitem{herrmann1979molecular}
A.~Herrmann, S.~Leutwyler, L.~W{\\"o}ste, and E.~Schumacher,
\newblock Chemical Physics Letters {\bf 62}, 444 (1979).

\bibitem{shuman2010laser}
E.~Shuman, J.~Barry, and D.~DeMille,
\newblock Nature , 820 (2010).

\bibitem{1972JChPh..57.1487D}
P.~J. {Dagdigian} and L.~{Wharton},
\newblock \jcp {\bf 57}, 1487 (1972).

\bibitem{2001PhRvA..64f3410B}
V.~I. {Balykin} and V.~S. {Letokhov},
\newblock \pra {\bf 64}, 063410 (2001).

\bibitem{2008PhRvL.100x0407T}
J.~J. {Thorn}, E.~A. {Schoene}, T.~{Li}, and D.~A. {Steck},
\newblock Physical Review Letters {\bf 100}, 240407 (2008), 0802.1585.

\bibitem{2011PhRvL.106p3002F}
M.~{Falkenau}, V.~V. {Volchkov}, J.~{R{\"u}hrig}, A.~{Griesmaier}, and
  T.~{Pfau},
\newblock Physical Review Letters {\bf 106}, 163002 (2011), 1102.0928.

\bibitem{2012PhRvA..85e3406S}
Y.~{Shagam} and E.~{Narevicius},
\newblock \pra {\bf 85}, 053406 (2012), 1112.0916.

\bibitem{2009NJPh...11e5046N}
E.~{Narevicius}, S.~T. {Bannerman}, and M.~G. {Raizen},
\newblock New Journal of Physics {\bf 11}, 055046 (2009), 0808.1383.

\bibitem{2010PhRvA..82b3419S}
E.~A. {Schoene}, J.~J. {Thorn}, and D.~A. {Steck},
\newblock \pra {\bf 82}, 023419 (2010).

\bibitem{1985JOSAB...2.1707D}
J.~{Dalibard} and C.~{Cohen-Tannoudji},
\newblock Journal of the Optical Society of America B Optical Physics {\bf 2},
  1707 (1985).

\bibitem{1983PhRvL..51.1336P}
D.~E. {Pritchard},
\newblock Physical Review Letters {\bf 51}, 1336 (1983).

\bibitem{1986PhRvL..57.1688A}
A.~{Aspect}, J.~{Dalibard}, A.~{Heidmann}, C.~{Salomon}, and
  C.~{Cohen-Tannoudji},
\newblock Physical Review Letters {\bf 57}, 1688 (1986).

\bibitem{1995PhRvL..74.2196N}
N.~R. {Newbury}, C.~J. {Myatt}, E.~A. {Cornell}, and C.~E. {Wieman},
\newblock Physical Review Letters {\bf 74}, 2196 (1995).

\bibitem{1995OptCo.119..652S}
J.~{S{\"o}ding}, R.~{Grimm}, and Y.~B. {Ovchinnikov},
\newblock Optics Communications {\bf 119}, 652 (1995).

\bibitem{2002PhRvA..66b3406M}
K.~W. {Miller}, S.~{D{\"u}rr}, and C.~E. {Wieman},
\newblock \pra {\bf 66}, 023406 (2002).

\bibitem{2005PhRvA..71a3422J}
J.~{Janis}, M.~{Banks}, and N.~P. {Bigelow},
\newblock \pra {\bf 71}, 013422 (2005).

\bibitem{2012Natur.491..570Z}
M.~{Zeppenfeld} {\em et~al.},
\newblock \nat {\bf 491}, 570 (2012), 1208.0046.

\bibitem{2009PhRvA..80d1401Z}
M.~{Zeppenfeld}, M.~{Motsch}, P.~W.~H. {Pinkse}, and G.~{Rempe},
\newblock \pra {\bf 80}, 041401 (2009), 0904.4144.

\bibitem{2009JPhB...42s5301R}
F.~{Robicheaux},
\newblock Journal of Physics B Atomic Molecular Physics {\bf 42}, 195301
  (2009).

\bibitem{1999PhRvL..83.1558B}
H.~L. {Bethlem}, G.~{Berden}, and G.~{Meijer},
\newblock \prl {\bf 83}, 1558 (1999).

\bibitem{2011PhRvL.106u3001W}
S.~{Wu}, R.~C. {Brown}, W.~D. {Phillips}, and J.~V. {Porto},
\newblock Physical Review Letters {\bf 106}, 213001 (2011), 1101.3602.

\bibitem{2011PhRvA..84f3417I}
V.~V. {Ivanov} and S.~{Gupta},
\newblock \pra {\bf 84}, 063417 (2011), 1110.3439.

\bibitem{2009ApPhB..94...71C}
L.~{Cabaret},
\newblock Applied Physics B: Lasers and Optics {\bf 94}, 71 (2009).

\bibitem{1994OptCo.106..202H}
B.~{Hoeling} and R.~J. {Knize},
\newblock Optics Communications {\bf 106}, 202 (1994).

\bibitem{2006PhRvL..97u3001P}
T.~{Pohl}, H.~R. {Sadeghpour}, Y.~{Nagata}, and Y.~{Yamazaki},
\newblock Physical Review Letters {\bf 97}, 213001 (2006).

\bibitem{2006JPhB...39.4945T}
C.~L. {Taylor}, J.~{Zhang}, and F.~{Robicheaux},
\newblock Journal of Physics B Atomic Molecular Physics {\bf 39}, 4945 (2006).

\bibitem{1989JOSAB...6.2023D}
J.~{Dalibard} and C.~{Cohen-Tannoudji},
\newblock Journal of the Optical Society of America B Optical Physics {\bf 6},
  2023 (1989).

\bibitem{ISI:000085557900004}
R.~Grimm, M.~Weidemuller, and Y.~Ovchinnikov,
\newblock ADVANCES IN ATOMIC MOLECULAR, AND OPTICAL PHYSICS, VOL. 42 {\bf 42},
  95 (2000).

\bibitem{Ketterle1996b}
W.~Ketterle and N.~J. van Druten,
\newblock Advances in Atomic, Molecular and Optical Physics {\bf 37}, 181
  (1996).

\bibitem{2006PhRvA..73d3410C}
D.~{Comparat} {\em et~al.},
\newblock Physical Review A {\bf 73}, 043410 (2006), quant-ph/0602010.

\bibitem{2009NJPh...11f3044T}
S.~{Travis Bannerman}, G.~N. {Price}, K.~{Viering}, and M.~G. {Raizen},
\newblock New Journal of Physics {\bf 11}, 063044 (2009).

\bibitem{Ketterle1996}
W.~Ketterle and N.~J. van Druten,
\newblock Phys. Rev. A {\bf 54}, 656 (1996).

\bibitem{1997PhRvA..55.3797S}
C.~A. {Sackett}, C.~C. {Bradley}, and R.~G. {Hulet},
\newblock \pra {\bf 55}, 3797 (1997).

\bibitem{2011arXiv1112.0916S}
Y.~{Shagam} and E.~{Narevicius},
\newblock ArXiv e-prints  (2011), 1112.0916.

\bibitem{2012arXiv1204.0352S}
V.~P. {Singh} and A.~{Ruschhaupt},
\newblock ArXiv e-prints  (2012), 1204.0352.

\bibitem{2011EPJD...65..161R}
J.~{Riedel} {\em et~al.},
\newblock European Physical Journal D {\bf 65}, 161 (2011).

\bibitem{2007PhRvA..76f3408H}
S.~{Hoekstra} {\em et~al.},
\newblock \pra {\bf 76}, 063408 (2007), arXiv:0709.3212.

\bibitem{2010NJPh...12f5028T}
E.~{Tsikata}, W.~C. {Campbell}, M.~T. {Hummon}, H.-I. {Lu}, and J.~M. {Doyle},
\newblock New Journal of Physics {\bf 12}, 065028 (2010).

\bibitem{2010JMoSp.260..115R}
R.~S. {Ram} and P.~F. {Bernath},
\newblock Journal of Molecular Spectroscopy {\bf 260}, 115 (2010).

\bibitem{2008PhRvL.101x3002S}
B.~K. {Stuhl}, B.~C. {Sawyer}, D.~{Wang}, and J.~{Ye},
\newblock Physical Review Letters {\bf 101}, 243002 (2008), 0808.2171.

\bibitem{1997PhRvA..55.4621O}
A.~M.~L. {Oien}, I.~T. {McKinnie}, P.~J. {Manson}, W.~J. {Sandle}, and D.~M.
  {Warrington},
\newblock \pra {\bf 55}, 4621 (1997).

\bibitem{Woh1}
W.~Wohlleben, F.~Chevy, K.~Madison, and J.~Dalibard,
\newblock Eur. Phys. J. D {\bf 15}, 237 (2001).

\bibitem{2008NJPh...10d5031C}
A.~{Chotia}, M.~{Viteau}, T.~{Vogt}, D.~{Comparat}, and P.~{Pillet},
\newblock New Journal of Physics {\bf 10}, 5031 (2008), arXiv:0803.4481.

\bibitem{1993PhRvL..70.2257S}
I.~D. {Setija} {\em et~al.},
\newblock Physical Review Letters {\bf 70}, 2257 (1993).

\bibitem{1995PhRvA..51...22T}
J.~J. {Tollett}, C.~C. {Bradley}, C.~A. {Sackett}, and R.~G. {Hulet},
\newblock \pra {\bf 51}, 22 (1995).

\bibitem{1995PhRvA..52.4004W}
J.~D. {Weinstein} and K.~G. {Libbrecht},
\newblock \pra {\bf 52}, 4004 (1995).

\bibitem{2005PhRvL..94g3003G}
J.~R. {Guest}, J.-H. {Choi}, E.~{Hansis}, A.~P. {Povilus}, and G.~{Raithel},
\newblock Physical Review Letters {\bf 94}, 073003 (2005).

\bibitem{2006ApPhB..82..533M}
K.~L. {Moore} {\em et~al.},
\newblock Applied Physics B: Lasers and Optics {\bf 82}, 533 (2006),
  arXiv:cond-mat/0504010.

\bibitem{MatthieuViteau07112008}
M.~Viteau {\em et~al.},
\newblock Science {\bf 321}, 232 (2008),
  http://www.sciencemag.org/cgi/reprint/321/5886/232.pdf.

\bibitem{2010NatPh...6..271S}
P.~F. {Staanum}, K.~{H{\o}jbjerre}, P.~S. {Skyt}, A.~K. {Hansen}, and
  M.~{Drewsen},
\newblock Nature Physics {\bf 6}, 271 (2010).

\bibitem{2010NatPh...6..275S}
T.~{Schneider}, B.~{Roth}, H.~{Duncker}, I.~{Ernsting}, and S.~{Schiller},
\newblock Nature Physics {\bf 6}, 275 (2010).

\bibitem{PhysRevLett.109.183001}
I.~Manai {\em et~al.},
\newblock Phys. Rev. Lett. {\bf 109}, 183001 (2012).

\bibitem{2009PhRvA..80e1401S}
D.~{Sofikitis} {\em et~al.},
\newblock \pra {\bf 80}, 051401 (2009).

\bibitem{2009NJPh...11e5037S}
D.~{Sofikitis} {\em et~al.},
\newblock New Journal of Physics {\bf 11}, 055037 (2009), 0903.1222.

\bibitem{2010NaPho...4..760C}
S.~T. {Cundiff} and A.~M. {Weiner},
\newblock Nature Photonics {\bf 4}, 760 (2010).

\bibitem{2008PhRvA..77b3402L}
B.~L. {Lev} {\em et~al.},
\newblock \pra {\bf 77}, 023402 (2008), 0705.3639.

\bibitem{1997PhRvL..78.1420S}
J.~{S{\"o}ding}, R.~{Grimm}, Y.~B. {Ovchinnikov}, P.~{Bouyer}, and
  C.~{Salomon},
\newblock Physical Review Letters {\bf 78}, 1420 (1997).

\bibitem{2009PhRvA..79f1407H}
E.~R. {Hudson},
\newblock \pra {\bf 79}, 061407 (2009), 0901.4164.

\bibitem{2010PhRvA..82e3408O}
C.~H.~R. {Ooi},
\newblock \pra {\bf 82}, 053408 (2010).

\bibitem{2011PhRvA..84f3401C}
M.~A. {Chieda} and E.~E. {Eyler},
\newblock \pra {\bf 84}, 063401 (2011), 1108.3543.

\bibitem{2012arXiv1201.1015I}
E.~{Ilinova}, J.~{Weinstein}, and A.~{Derevianko},
\newblock ArXiv e-prints  (2012), 1201.1015.

\bibitem{2008PhRvL.100h3003C}
W.~C. {Campbell}, G.~C. {Groenenboom}, H.-I. {Lu}, E.~{Tsikata}, and J.~M.
  {Doyle},
\newblock Physical Review Letters {\bf 100}, 083003 (2008), 0711.0379.

\bibitem{2005PhRvL..95a3003V}
S.~Y. {van de Meerakker}, N.~{Vanhaecke}, M.~P. {van der Loo}, G.~C.
  {Groenenboom}, and G.~{Meijer},
\newblock Physical Review Letters {\bf 95}, 013003 (2005),
  arXiv:physics/0505097.

\bibitem{2010JOpt...12k3001T}
J.~C. {Travers},
\newblock Journal of Optics {\bf 12}, 113001 (2010).

\bibitem{2011OptCo.284.3669W}
A.~M. {Weiner},
\newblock Optics Communications {\bf 284}, 3669 (2011).

\bibitem{Wilson:07}
J.~W. Wilson, P.~Schlup, and R.~A. Bartels,
\newblock Opt. Express {\bf 15}, 8979 (2007).

\bibitem{Frumker:07}
E.~Frumker and Y.~Silberberg,
\newblock Opt. Lett. {\bf 32}, 1384 (2007).

\bibitem{Fontaine:08}
N.~K. Fontaine {\em et~al.},
\newblock Opt. Lett. {\bf 33}, 1714 (2008).

\bibitem{Shirasaki:96}
M.~Shirasaki,
\newblock Opt. Lett. {\bf 21}, 366 (1996).

\bibitem{Xiao:05}
S.~Xiao, A.~M. Weiner, and C.~Lin,
\newblock J. Lightwave Technol. {\bf 23}, 1456 (2005).

\bibitem{Willits:12}
J.~T. Willits, A.~M. Weiner, and S.~T. Cundiff,
\newblock Opt. Express {\bf 20}, 3110 (2012).

\bibitem{Supradeepa:08}
V.~R. Supradeepa, C.-B. Huang, D.~E. Leaird, and A.~M. Weiner,
\newblock Opt. Express {\bf 16}, 11878 (2008).

\bibitem{2004EPJD...31..395D}
M.~D. {di Rosa},
\newblock European Physical Journal D {\bf 31}, 395 (2004).

\bibitem{2009PhRvL.103v3001S}
E.~S. {Shuman}, J.~F. {Barry}, D.~R. {Glenn}, and D.~{Demille},
\newblock Physical Review Letters {\bf 103}, 223001 (2009), 0909.2600.

\bibitem{2010PhRvA..82e2521I}
T.~A. {Isaev}, S.~{Hoekstra}, and R.~{Berger},
\newblock \pra {\bf 82}, 052521 (2010), 1007.1788.

\bibitem{2011PhRvA..83e3404N}
J.~H.~V. {Nguyen} and B.~{Odom},
\newblock \pra {\bf 83}, 053404 (2011), 1012.3696.

\bibitem{2011NJPh...13f3023N}
J.~H.~V. {Nguyen} {\em et~al.},
\newblock New Journal of Physics {\bf 13}, 063023 (2011), 1102.3368.

\bibitem{2010PhRvA..82c3415C}
S.~{Choi}, B.~{Sundaram}, and M.~G. {Raizen},
\newblock \pra {\bf 82}, 033415 (2010), 1008.1255.

\bibitem{1998PhRvL..81.2194S}
D.~M. {Stamper-Kurn} {\em et~al.},
\newblock Physical Review Letters {\bf 81}, 2194 (1998).

\bibitem{varshalovich1987quantum}
D.~Varshalovich, A.~Moskalev, and V.~Khersonskii,
\newblock {\em {Quantum theory of angular momentum}} (World Scientific Pub.
  Co., Teaneck, NJ, 1987).

\bibitem{BrownCarrington2003}
J.~Brown and A.~Carrington,
\newblock {\em Rotational Spectroscopy of Diatomic Molecules} (Cambridge
  University Press, 2003).

\bibitem{CDG2}
C.~Cohen-Tannoudji, J.~Dupond-Roc, and G.~Grynberg,
\newblock {\em Processus d'Interaction entre Photons et Atomes} (InterEdition,
  Paris, 1988).

\bibitem{2012AmJPh..80..882H}
R.~{H{\"o}ppner}, E.~{Rold{\'a}n}, and G.~J. {de Valc{\'a}rcel},
\newblock American Journal of Physics {\bf 80}, 882 (2012), 1105.4120.

\bibitem{2004PhRvA..69f3806B}
K.~{Blushs} and M.~{Auzinsh},
\newblock \pra {\bf 69}, 063806 (2004), arXiv:physics/0312124.

\bibitem{2003PhRvA..67b2902M}
T.~{Minami}, C.~O. {Reinhold}, and J.~{Burgd{\"o}rfer},
\newblock \pra {\bf 67}, 022902 (2003).

\bibitem{2010PhRvA..82a3615G}
F.~{Gerbier} and Y.~{Castin},
\newblock \pra {\bf 82}, 013615 (2010), 1002.5018.

\bibitem{Gri1}
R.~Grimm, M.~Weidemï¿½ller, and Y.~Ovchinnikov,
\newblock Adv. At. Mol. Opt. Phys. {\bf 42}, 95 (2000),
\newblock arXiv:physics/9902072.

\bibitem{2003cond.mat..3028J}
A.~P.~J. {Jansen},
\newblock eprint arXiv:cond-mat/0303028  (2003), arXiv:cond-mat/0303028.

\bibitem{heggie2003gravitational}
D.~Heggie and P.~Hut,
\newblock {\em The gravitational million-body problem} (Cambridge University
  Press, 2003).

\bibitem{Aarseth}
S.~J. Aarseth,
\newblock {\em Gravitational N-body Simulations: Tools and Algorithms}
  (Cambridge Monographs on Mathematical Physics, 2006).

\bibitem{2008NewA...13..285P}
S.~{Portegies Zwart} {\em et~al.},
\newblock New Astronomy {\bf 13}, 285 (2008), 0711.0643.

\bibitem{1990PhLA..150..262Y}
H.~{Yoshida},
\newblock Physics Letters A {\bf 150}, 262 (1990).

\bibitem{2005PhRvL..94l3201C}
C.~{Chin} {\em et~al.},
\newblock Physical Review Letters {\bf 94}, 123201 (2005),
  arXiv:cond-mat/0411258.

\bibitem{2004PhRvL..93l3202T}
Y.~{Takasu} {\em et~al.},
\newblock Physical Review Letters {\bf 93}, 123202 (2004).

\bibitem{2010RvMP...82.1225C}
C.~{Chin}, R.~{Grimm}, P.~{Julienne}, and E.~{Tiesinga},
\newblock Reviews of Modern Physics {\bf 82}, 1225 (2010).

\bibitem{2006RvMP...78.1311K}
T.~{K{\"o}hler}, K.~{G{\'o}ral}, and P.~S. {Julienne},
\newblock Reviews of Modern Physics {\bf 78}, 1311 (2006),
  arXiv:cond-mat/0601420.

\bibitem{2002Natur.417..529D}
E.~A. {Donley}, N.~R. {Claussen}, S.~T. {Thompson}, and C.~E. {Wieman},
\newblock \nat {\bf 417}, 529 (2002), arXiv:cond-mat/0204436.

\bibitem{2005PhRvL..95s0404T}
S.~T. {Thompson}, E.~{Hodby}, and C.~E. {Wieman},
\newblock Physical Review Letters {\bf 95}, 190404 (2005),
  arXiv:cond-mat/0505567.

\bibitem{2006PhRvL..97r0404P}
S.~B. {Papp} and C.~E. {Wieman},
\newblock Physical Review Letters {\bf 97}, 180404 (2006),
  arXiv:cond-mat/0607667.

\bibitem{2003Sci...301.1510H}
J.~{Herbig} {\em et~al.},
\newblock Science {\bf 301}, 1510 (2003).

\bibitem{regal2003}
C.~A. Regal, C.~Ticknor, J.~L. Bohn, and D.~S. Jin,
\newblock Nature {\bf 424}, 47 (2003).

\bibitem{2003Sci...302.2101J}
S.~{Jochim} {\em et~al.},
\newblock Science {\bf 302}, 2101 (2003).

\bibitem{2003PhRvL..91x0401C}
J.~{Cubizolles}, T.~{Bourdel}, S.~J. {Kokkelmans}, G.~V. {Shlyapnikov}, and
  C.~{Salomon},
\newblock Physical Review Letters {\bf 91}, 240401 (2003),
  arXiv:cond-mat/0308018.

\bibitem{2003PhRvL..91h0406S}
K.~E. {Strecker}, G.~B. {Partridge}, and R.~G. {Hulet},
\newblock Physical Review Letters {\bf 91}, 080406 (2003),
  arXiv:cond-mat/0308318.

\bibitem{xu2003formation}
K.~Xu {\em et~al.},
\newblock Physical review letters {\bf 91}, 210402 (2003).

\bibitem{2006PhRvL..97l0402O}
C.~{Ospelkaus} {\em et~al.},
\newblock Physical Review Letters {\bf 97}, 120402 (2006),
  arXiv:cond-mat/0607581.

\bibitem{2008PhRvA..78f1601W}
C.~{Weber} {\em et~al.},
\newblock \pra {\bf 78}, 061601 (2008), 0808.4077.

\bibitem{2009PhRvA..79c2706B}
Q.~{Beaufils} {\em et~al.},
\newblock \pra {\bf 79}, 032706 (2009), 0811.4282.

\bibitem{2010Sci...330..940L}
T.~{Lompe} {\em et~al.},
\newblock Science {\bf 330}, 940 (2010), 1006.2241.

\bibitem{2012PhRvL.109h5301W}
C.-H. {Wu}, J.~W. {Park}, P.~{Ahmadi}, S.~{Will}, and M.~W. {Zwierlein},
\newblock Physical Review Letters {\bf 109}, 085301 (2012), 1206.5023.

\bibitem{2006RvMP...78..483J}
K.~M. {Jones}, E.~{Tiesinga}, P.~D. {Lett}, and P.~S. {Julienne},
\newblock Reviews of Modern Physics {\bf 78}, 483 (2006).

\bibitem{1998PhRvL..80.4402F}
A.~{Fioretti} {\em et~al.},
\newblock \prl {\bf 80}, 4402 (1998).

\bibitem{1999PhRvL..82..307M}
A.~P. {Mosk}, M.~W. {Reynolds}, T.~W. {Hijmans}, and J.~T.~M. {Walraven},
\newblock Physical Review Letters {\bf 82}, 307 (1999).

\bibitem{2000PhRvL..84.2814G}
C.~{Gabbanini}, A.~{Fioretti}, A.~{Lucchesini}, S.~{Gozzini}, and M.~{Mazzoni},
\newblock Physical Review Letters {\bf 84}, 2814 (2000).

\bibitem{2000Natur.408..692G}
J.~M. {Gerton}, D.~{Strekalov}, I.~{Prodan}, and R.~G. {Hulet},
\newblock \nat {\bf 408}, 692 (2000), arXiv:cond-mat/0011513.

\bibitem{2002PhRvA..66e3401F}
F.~K. {Fatemi}, K.~M. {Jones}, P.~D. {Lett}, and E.~{Tiesinga},
\newblock \pra {\bf 66}, 053401 (2002).

\bibitem{nikolov2000}
A.~N. Nikolov {\em et~al.},
\newblock Phys. Rev. Lett. {\bf 84}, 246 (2000).

\bibitem{2003Natur.426..537G}
M.~{Greiner}, C.~A. {Regal}, and D.~S. {Jin},
\newblock \nat {\bf 426}, 537 (2003).

\bibitem{2007PhRvL..98t0403G}
J.~P. {Gaebler}, J.~T. {Stewart}, J.~L. {Bohn}, and D.~S. {Jin},
\newblock Physical Review Letters {\bf 98}, 200403 (2007),
  arXiv:cond-mat/0703087.

\bibitem{2006PhRvL..96b3203M}
S.~{Moal} {\em et~al.},
\newblock Physical Review Letters {\bf 96}, 023203 (2006),
  arXiv:cond-mat/0509286.

\bibitem{2000PhRvL..85.2292Z}
G.~{Zinner}, T.~{Binnewies}, F.~{Riehle}, and E.~{Tiemann},
\newblock Physical Review Letters {\bf 85}, 2292 (2000).

\bibitem{2004PhRvL..93x3005W}
D.~{Wang} {\em et~al.},
\newblock Physical Review Letters {\bf 93}, 243005 (2004),
  arXiv:physics/0410220.

\bibitem{2004PhRvL..92o3001K}
A.~J. {Kerman}, J.~M. {Sage}, S.~{Sainis}, T.~{Bergeman}, and D.~{Demille},
\newblock Physical Review Letters {\bf 92}, 153001 (2004),
  arXiv:physics/0402116.

\bibitem{2004PhRvA..70b1402H}
C.~{Haimberger}, J.~{Kleinert}, M.~{Bhattacharya}, and N.~P. {Bigelow},
\newblock \pra {\bf 70}, 021402 (2004).

\bibitem{2008PhRvL.101m3004D}
J.~{Deiglmayr} {\em et~al.},
\newblock Physical Review Letters {\bf 101}, 133004 (2008), 0807.3272.

\bibitem{2003PhRvL..91y0401Z}
M.~W. {Zwierlein} {\em et~al.},
\newblock Physical Review Letters {\bf 91}, 250401 (2003),
  arXiv:cond-mat/0311617.

\bibitem{2013PhRvL.110n3001H}
M.~T. {Hummon} {\em et~al.},
\newblock Physical Review Letters {\bf 110}, 143001 (2013), 1209.4069.

\bibitem{2013arXiv1308.0421Z}
V.~{Zhelyazkova} {\em et~al.},
\newblock ArXiv e-prints  (2013), 1308.0421.

\bibitem{2011arXiv1111.2841H}
N.~R. {Hutzler}, H.-I. {Lu}, and J.~M. {Doyle},
\newblock ArXiv e-prints  (2011), 1111.2841.

\bibitem{1998Natur.395..148W}
J.~D. {Weinstein}, R.~{Decarvalho}, T.~{Guillet}, B.~{Friedrich}, and J.~M.
  {Doyle},
\newblock \nat {\bf 395}, 148 (1998).

\bibitem{1998JChPh.109.2656W}
J.~D. {Weinstein} {\em et~al.},
\newblock \jcp {\bf 109}, 2656 (1998).

\bibitem{2005PhRvL..94l3002M}
K.~{Maussang}, D.~{Egorov}, J.~S. {Helton}, S.~V. {Nguyen}, and J.~M. {Doyle},
\newblock Physical Review Letters {\bf 94}, 123002 (2005).

\bibitem{2001PhRvA..63c0501E}
D.~{Egorov}, J.~D. {Weinstein}, D.~{Patterson}, B.~{Friedrich}, and J.~M.
  {Doyle},
\newblock \pra {\bf 63}, 030501 (2001).

\bibitem{2005PhRvL..95q3201M}
S.~E. {Maxwell} {\em et~al.},
\newblock Physical Review Letters {\bf 95}, 173201 (2005),
  arXiv:physics/0508100.

\bibitem{2007JChPh.126o4307P}
D.~{Patterson} and J.~M. {Doyle},
\newblock \jcp {\bf 126}, 4307 (2007).

\bibitem{2004EPJD...31..307E}
D.~{Egorov} {\em et~al.},
\newblock European Physical Journal D {\bf 31}, 307 (2004).

\bibitem{2008PhRvA..78c2707S}
M.~{Stoll}, J.~M. {Bakker}, T.~C. {Steimle}, G.~{Meijer}, and A.~{Peters},
\newblock \pra {\bf 78}, 032707 (2008).

\bibitem{2009PhRvL.102c3001V}
L.~D. {van Buuren} {\em et~al.},
\newblock Physical Review Letters {\bf 102}, 033001 (2009), 0806.2523.

\bibitem{2009NJPh...11l3026S}
S.~M. {Skoff} {\em et~al.},
\newblock New Journal of Physics {\bf 11}, 123026 (2009), 0909.5534.

\bibitem{2010APS..MARH24001D}
J.~{Doyle},
\newblock {Cold, Trapped Molecules via Cryogenic Buffer Gas Methods},
\newblock in {\em APS Meeting Abstracts}, p. 24001, 2010.

\bibitem{2008NatPh...4..595V}
S.~Y.~T. {van de Meerakker}, H.~L. {Bethlem}, and G.~{Meijer},
\newblock Nature Physics {\bf 4}, 595 (2008).

\bibitem{2000Natur.406..491B}
H.~L. {Bethlem} {\em et~al.},
\newblock \nat {\bf 406}, 491 (2000).

\bibitem{2001Natur.411..174C}
F.~M.~H. {Crompvoets}, H.~L. {Bethlem}, R.~T. {Jongma}, and G.~{Meijer},
\newblock \nat {\bf 411}, 174 (2001).

\bibitem{2003PhRvL..91x3001B}
J.~R. {Bochinski}, E.~R. {Hudson}, H.~J. {Lewandowski}, G.~{Meijer}, and
  J.~{Ye},
\newblock Physical Review Letters {\bf 91}, 243001 (2003),
  arXiv:physics/0306062.

\bibitem{2004PhRvA..70d3410B}
J.~R. {Bochinski}, E.~R. {Hudson}, H.~J. {Lewandowski}, and J.~{Ye},
\newblock \pra {\bf 70}, 043410 (2004), arXiv:physics/0403126.

\bibitem{2005PhRvL..94b3004V}
S.~Y. {van de Meerakker}, P.~H. {Smeets}, N.~{Vanhaecke}, R.~T. {Jongma}, and
  G.~{Meijer},
\newblock Physical Review Letters {\bf 94}, 023004 (2005),
  arXiv:physics/0407116.

\bibitem{2006PhRvA..73f3404H}
E.~R. {Hudson} {\em et~al.},
\newblock \pra {\bf 73}, 063404 (2006), arXiv:physics/0508120.

\bibitem{2006JPhB...39.1077V}
S.~Y.~T. {van de Meerakker}, I.~{Labazan}, S.~{Hoekstra}, J.~{K{\"u}pper}, and
  G.~{Meijer},
\newblock Journal of Physics B Atomic Molecular Physics {\bf 39}, 1077 (2006),
  arXiv:physics/0512194.

\bibitem{2008EPJD...46..463B}
O.~{Bucicov} {\em et~al.},
\newblock European Physical Journal D {\bf 46}, 463 (2008), 0712.1526.

\bibitem{2008PhRvA..77c1404W}
K.~{Wohlfart} {\em et~al.},
\newblock \pra {\bf 77}, 031404 (2008), arXiv:0803.0650.

\bibitem{2004PhRvL..92q3002T}
M.~R. {Tarbutt} {\em et~al.},
\newblock Physical Review Letters {\bf 92}, 173002 (2004),
  arXiv:physics/0312119.

\bibitem{2009NJPh...11e5038T}
S.~K. {Tokunaga}, J.~M. {Dyne}, E.~A. {Hinds}, and M.~R. {Tarbutt},
\newblock New Journal of Physics {\bf 11}, 055038 (2009), 0812.4188.

\bibitem{2010PhRvA..81c3414W}
T.~E. {Wall}, S.~K. {Tokunaga}, E.~A. {Hinds}, and M.~R. {Tarbutt},
\newblock \pra {\bf 81}, 033414 (2010), 1001.3222.

\bibitem{2004PhRvL..92c3005V}
E.~{Vliegen}, H.~J. {W{\"o}rner}, T.~P. {Softley}, and F.~{Merkt},
\newblock Physical Review Letters {\bf 92}, 033005 (2004).

\bibitem{2007JPhCS..80a2045Y}
Y.~{Yamakita} {\em et~al.},
\newblock Journal of Physics Conference Series {\bf 80}, 2045 (2007).

\bibitem{2008PhRvL.100d3001H}
S.~D. {Hogan} and F.~{Merkt},
\newblock Physical Review Letters {\bf 100}, 043001 (2008).

\bibitem{2009PhRvL.103l3001H}
S.~D. {Hogan}, C.~{Seiler}, and F.~{Merkt},
\newblock Physical Review Letters {\bf 103}, 123001 (2009).

\bibitem{2004PhRvL..93x3004F}
R.~{Fulton}, A.~I. {Bishop}, and P.~F. {Barker},
\newblock \prl {\bf 93}, 243004 (2004).

\bibitem{2006JPhB...39.1097F}
R.~{Fulton}, A.~I. {Bishop}, M.~N. {Shneider}, and P.~F. {Barker},
\newblock Journal of Physics B Atomic Molecular Physics {\bf 39}, 1097 (2006).

\bibitem{2008PhRvA..77e1401N}
E.~{Narevicius} {\em et~al.},
\newblock \pra {\bf 77}, 051401 (2008).

\bibitem{2003Sci...302.1940E}
M.~S. {Elioff}, J.~J. {Valentini}, and D.~W. {Chandler},
\newblock Science {\bf 302}, 1940 (2003).

\bibitem{2007PhRvL..98j3002L}
N.-N. {Liu} and H.~{Loesch},
\newblock Physical Review Letters {\bf 98}, 103002 (2007).

\bibitem{2009FaDi..142..143K}
J.~J. {Kay}, S.~Y.~T. {van de Meerakker}, K.~E. {Strecker}, and D.~W.
  {Chandler},
\newblock Cold and Ultracold Molecules, Faraday Discussions, volume 142, 2009,
  p.143 {\bf 142}, 143 (2009).

\bibitem{2010arXiv1002.3698T}
A.~{Trottier}, D.~{Carty}, and E.~{Wrede},
\newblock ArXiv e-prints  (2010), 1002.3698.

\bibitem{gupta1999}
M.~Gupta and D.~Herschbach,
\newblock J. Phys. Chem. A {\bf 103}, 10670 (1999).

\bibitem{gupta2001}
M.~Gupta and D.~Herschbach,
\newblock J. Phys. Chem. A {\bf 105}, 1626 (2001).

\bibitem{2010PhRvA..81c3409S}
M.~{Strebel}, F.~{Stienkemeier}, and M.~{Mudrich},
\newblock \pra {\bf 81}, 033409 (2010), 0910.3278.

\bibitem{2008EPJD...46..307D}
S.~{Deachapunya} {\em et~al.},
\newblock European Physical Journal D {\bf 46}, 307 (2008), arXiv:0708.1449.

\bibitem{2003PhRvA..67d3406R}
S.~A. {Rangwala}, T.~{Junglen}, T.~{Rieger}, P.~W. {Pinkse}, and G.~{Rempe},
\newblock \pra {\bf 67}, 043406 (2003), arXiv:physics/0209041.

\bibitem{2006PhRvA..73f1402R}
T.~{Rieger} {\em et~al.},
\newblock \pra {\bf 73}, 061402 (2006), arXiv:physics/0512119.

\bibitem{2008PhRvL.100d3203W}
S.~{Willitsch}, M.~T. {Bell}, A.~D. {Gingell}, S.~R. {Procter}, and T.~P.
  {Softley},
\newblock Physical Review Letters {\bf 100}, 043203 (2008).

\bibitem{2009FaDi..142..203S}
C.~{Sommer} {\em et~al.},
\newblock Faraday Discussions {\bf 142}, 203 (2009), 0812.1923.

\bibitem{2010PCCP...12..745L}
Y.~{Liu}, M.~{Yun}, Y.~{Xia}, L.~{Deng}, and J.~{Yin},
\newblock Physical Chemistry Chemical Physics (Incorporating Faraday
  Transactions) {\bf 12}, 745 (2010).

\bibitem{2009PhRvA..79a3405M}
M.~{Motsch} {\em et~al.},
\newblock \pra {\bf 79}, 013405 (2009), 0809.1728.

\bibitem{2010JPhB...43i5202T}
H.~{Tsuji}, T.~{Sekiguchi}, T.~{Mori}, T.~{Momose}, and H.~{Kanamori},
\newblock Journal of Physics B Atomic Molecular Physics {\bf 43}, 095202
  (2010).

\bibitem{2009JPhB...42o4001W}
R.~{Wester},
\newblock Journal of Physics B Atomic Molecular Physics {\bf 42}, 154001
  (2009), 0902.0475.

\bibitem{low_temp_Mol_2008}
I.~W.~M. Smith, editor,
\newblock {\em Low Temperatures And Cold Molecules} (Imperial College Press,
  2008).

\bibitem{cold_Mol_2009}
W.~C.~S. R.~V.~Krems and B.~Friedrich, editors,
\newblock {\em Cold molecules: theory, experiment, applications} (CRC Press,
  Boca Raton, Florida, 2009).

\bibitem{1997quant.ph.10025W}
D.~J. {Wineland} {\em et~al.},
\newblock eprint arXiv:quant-ph/9710025  (1997), arXiv:quant-ph/9710025.

\bibitem{2006PhRvL..97x3005O}
A.~{Ostendorf} {\em et~al.},
\newblock Physical Review Letters {\bf 97}, 243005 (2006).

\bibitem{phdHornekaer2000}
L.~Hornekï¿½r,
\newblock {\em Single- and Multi-Species Coulomb Ion Crystals: Structures,
  Dynamics and Sympathetic Cooling},
\newblock PhD thesis, Danish National Research Foundation Center for Quantum
  Optics - Quantop Department of Physics and Astronomy, The University of
  Aarhus, 2000.

\bibitem{2005PhRvL..95r3002B}
P.~{Blythe}, B.~{Roth}, U.~{Fr{\"o}hlich}, H.~{Wenz}, and S.~{Schiller},
\newblock Physical Review Letters {\bf 95}, 183002 (2005).

\bibitem{2008PhRvA..78d2709R}
B.~{Roth}, D.~{Offenberg}, C.~B. {Zhang}, and S.~{Schiller},
\newblock \pra {\bf 78}, 042709 (2008).

\bibitem{2008PhRvA..78f1401O}
D.~{Offenberg}, C.~B. {Zhang}, C.~{Wellers}, B.~{Roth}, and S.~{Schiller},
\newblock \pra {\bf 78}, 061401 (2008), 0810.5102.

\bibitem{2009JPhB...42c5101O}
D.~{Offenberg}, C.~{Wellers}, C.~B. {Zhang}, B.~{Roth}, and S.~{Schiller},
\newblock Journal of Physics B Atomic Molecular Physics {\bf 42}, 035101
  (2009), 0810.5097.

\bibitem{2000JChPh.112.8068E}
U.~{Even}, J.~{Jortner}, D.~{Noy}, N.~{Lavie}, and C.~{Cossart-Magos},
\newblock \jcp {\bf 112}, 8068 (2000).

\bibitem{choi2006infrared}
M.~Choi {\em et~al.},
\newblock International Reviews in Physical Chemistry {\bf 25}, 15 (2006).

\bibitem{2006JPhB...39R.127S}
F.~{Stienkemeier} and K.~K. {Lehmann},
\newblock Journal of Physics B Atomic Molecular Physics {\bf 39}, 127 (2006),
  arXiv:physics/0604090.

\bibitem{2006JLTP..142....1B}
M.~{Barranco} {\em et~al.},
\newblock Journal of Low Temperature Physics {\bf 142}, 1 (2006).

\bibitem{2007IRPC...26..249K}
J.~{K{\"u}pper} and J.~M. {Merritt},
\newblock International Reviews in Physical Chemistry {\bf 26}, 249 (2007),
  arXiv:physics/0609052.

\bibitem{Handbook_spectro}
F.~M. Martin~Quack,
\newblock {\em Handbook of High-Resolution Spectroscopy} (John Wiley \& Sons,
  2011).

\bibitem{2010RScI...81e4101F}
T.~M. {Falconer}, W.~K. {Lewis}, R.~J. {Bemish}, R.~E. {Miller}, and G.~L.
  {Glish},
\newblock Review of Scientific Instruments {\bf 81}, 054101 (2010).

\end{thebibliography}

\end{document}